\newcommand{\cmthree}{\;{\rm cm$^{-3}$}}
\newcommand{\hers}{{\it  Herschel}}

\newcommand{\spitz}{{\it   Spitzer}}

\newcommand{\mic}{$\mu$m}

\newcommand{\lsol}{L$_\odot$}
\newcommand{\unitsurf}{W\,m$^{-2}$sr$^{-1}$}
\newcommand{\zsol}{Z$_\odot$}

\newcommand{\smallsub}[1]{{\mbox{{\tiny #1}}}}

\newcommand{\eg}{e.g.}

\newcommand{\Av}{A$_\smallsub{V}$}

\newcommand{\n}{$n_\smallsub{0}$} 

\newcommand{\chii}{c$_\mathrm{H\,{\sc II}}$}
\newcommand{\cpdr}{$c_{\rm PDR}$}
\newcommand{\specie}[2]{$[$#1{~\sc #2}$]$}
\newcommand{\PDRHIIratio}{(\specie{C}{ii}+\specie{O}{i})/\specie{O}{iii}}
\newcommand{\Rpdrhii}{R$_{\rm PDR/HII}$}
\newcommand{\OIIItfratio}{\specie{O}{iii}/24\mic}

\newcommand{\OIIIsvratio}{\specie{O}{iii}/70\mic}

\newcommand{\CIIPDRratio}{\specie{C}{ii}/(\specie{C}{ii}+\specie{O}{i})}
\newcommand{\Rciipdr}{R$_{\mathrm{C\,{\sc II}}/{\rm PDR}}$}

\newcommand{\co}{CO$\,$(1-0)}
\newcommand{\hmol}{H$_2$}

\newcommand{\HII}{H$\,${~\sc ii}}

\newcommand{\cplus}{C$^+$}
\newcommand{\oplusplus}{O$^{++}$}

\newcommand{\cii}{[C$\,${~\sc ii}]}

\newcommand{\oiii}{[O$\,${~\sc iii}]}
\newcommand{\oi}{[O$\,${~\sc i}]}

\newcommand{\ciiline}{\cii$\lambda 158$\mic}

\newcommand{\oiiiline}{\oiii$\lambda 88$\mic}
\newcommand{\oilinelo}{\oi$\lambda 63$\mic}

\newcommand{\loiii}{L$_\mathrm{[O\,{\sc III}]}$}
\newcommand{\lcont}{L$_\mathrm{[O\,{\sc 24}]}$}

\documentclass[runningheads,a4paper]{aa}  

\usepackage{graphicx}
\usepackage{txfonts}
\usepackage{xcolor}
\usepackage{longtable}
\usepackage{array}
\usepackage[section]{placeins}
\usepackage[flushleft]{threeparttable}
\usepackage{natbib}
\usepackage[breaklinks=true]{hyperref}
\bibpunct{(}{)}{;}{a}{}{,}
\hypersetup{colorlinks=true,linkcolor=magenta,citecolor=blue,filecolor=blue,urlcolor=magenta}

\begin{document} 
\newcolumntype{C}[1]{>{\centering\let\newline\\\arraybackslash\hspace{0pt}}m{#1}}

\graphicspath{{images/}}

\title{Disentangling emission from star-forming regions in the Magellanic Clouds: Linking \oiiiline\ and 24\mic}

   \author{A. Lambert-Huyghe\inst{1} 
           \and S. C. Madden\inst{1}
           \and V. Lebouteiller\inst{1}
           \and F. Galliano\inst{1}
           \and N. Abel\inst{2}
           \and D. Hu\inst{1}
           \and L. Ramambason\inst{1}
           \and F. L. Polles\inst{3}}

   \institute{AIM, CEA, CNRS, Universit\'e Paris-Saclay, Universit\'e Paris Diderot, Sorbonne Paris, 91191 Gif-sur-Yvette, France \email{antigonelh18@hotmail.fr} 
              \and University of Cincinnati - Clermont Campus, Batavia, OH, USA 
              \and SOFIA Science Center, USRA, NASA Ames Research Center, M.S. N232-12, Moffett Field, CA, 94035, USA }

   \date{Received 28/03/2022; accepted 21/06/2022}

 
  \abstract
   {The \oiiiline\ line is observed in many galaxies including our neighboring Magellanic Clouds and is a well-known tracer of \HII\ regions, while the 24\mic\ continuum emission has often been used to trace warm dust in the ionized phases of galaxies. The association of both the \oiiiline\ line and 24\mic\ in galaxies to star formation motivates this study to determine their observational relation. } 
   {This study explores the link between the \oiiiline\ and 24\mic\ continuum in star-forming regions in the Magellanic Clouds. We also explore the local conditions driving the relation between those tracers. }
   {We compared observations with 1D Cloudy models consisting of an \HII\ region plus a photodissociation region (PDR) component, varying the stellar age, the initial density (at the illuminated edge of the cloud), and the ionization parameter. We introduced a new parameter, \cpdr, to quantify the proportion of emission arising from PDRs and that with an origin in \HII\ regions along each line of sight. We used the ratio \PDRHIIratio\, as a proxy for the ratio of PDR versus \HII\ region emission, and compared it to the \OIIItfratio\ ratio. The use of \OIIItfratio\ and \OIIIsvratio\ together allowed us to constrain the models most efficiently.}
   {We find a correlation over at least 3 orders of magnitude in \oiiiline\ and 24\mic\ continuum. This correlation is seen for spatially resolved maps of the Magellanic Cloud regions as well as unresolved galaxy-wide low metallicity galaxies of the Dwarf Galaxy Survey. We also find that most of the regions have low proportions of PDRs along the lines of sight ($<$ 12\%), while a limited area of some of the mapped regions can reach 30 to 50\%. For most lines of sight within the star-forming regions we have studied in the Magellanic Clouds, \HII\ regions are the dominant phase.}
   {We propose the use of the correlation between the \oiiiline\ and 24\mic\ continuum as a new predictive tool to estimate, for example, the \oiiiline\ when the 24\mic\ continuum is available or inversely. This can be especially useful to prepare for {\it Atacama Large Milimeter Array} (ALMA) observations of  \oiiiline\ in high-z galaxies. The simple and novel method we developed may also provides a way to disentangle different phases along the line of sight, when other 3D information is not available.}

               
\authorrunning{A. Lambert-Huyghe et al.}
\maketitle
 %

\section{Introduction}
\label{intro}
The interstellar medium (ISM) harbors many processes that play key roles in the evolution of galaxies. It is composed of material that has been accreted from the intergalactic reservoir as well as the heavy elements (metals heavier than hydrogen or helium) produced in stars and returned to the ISM at the end of their lives to feed the next generation of stars. In return, this star formation process and the interaction between stars and the ISM evolve with time, as the maturing stars are cycling their newly made elements to the ISM, enriching the latter ISM reservoir with metals from each generation.  Despite recent progress in high-z surveys \citep[e.g.,][]{Maiolino15, Inoue16, Knudsen16, Pentericci16, Matthee17, Matthee19, Carniani18, Carniani20, Hashimoto19a, Hashimoto19b, Hashimoto19c, Hashimoto20, LeFevre20, Izumi21}, following the process of metal enrichment as galaxies evolve through time, from the earliest galaxies to the present day, has not been an easy observational endeavor with the current telescope sensitivities. Thus, many questions are still open regarding low metallicity ISM conditions and the formation of stars at the earliest epochs. A robust understanding of the origins of the tracers of star-forming regions and insight into their diagnostic capabilities are some of the necessary steps to guide the interpretation of high-z ISM and star-forming conditions.  

While not completely mimicking the earliest galaxies, local Universe dwarf galaxies are often used as laboratories to understand the physical properties of the gas and dust and their interplay with star formation in early Universe environments. The proximity of nearby dwarf galaxies makes many different tracers accessible, some of which are too faint to be detected in more distant galaxies. Moreover, the diverse collection of local Universe dwarf galaxies exhibit a wide range of metallicities and star formation rates (SFR). 

Local dwarf galaxies were the focus of large \hers\ and \spitz\ surveys \citep[e.g., The Dwarf Galaxy Survey, DGS;][] {Madden06}. Studies on both resolved  and integrated-galaxy scales have highlighted some distinctively unique observational signatures of star-forming low-metallicity dwarf galaxies. A nonlinear relation of the dust-to-gas mass (D/G) with metallicity is observed, with extremely low dust masses observed for the lowest metallicity galaxies (Z $\leq$ 0.1 \zsol) \citep{Herrera12, Fisher14, Remy15, Galliano18b, Galliano21, Cigan21}. Furthermore, the hard radiation fields in star-forming dwarf galaxies, along with their lower dust abundance, result in extended ionized gas regions prominent on global galaxy scales \citep{Hunter11, Cormier15, Cormier19}. The consequence is often a largely photodissociated molecular phase, existing in clumps which are difficult to observe with the usual molecular gas tracer, \co\ \citep{Cormier14, Hunt15, Accurso17b}, beckoning the presence of a CO-dark molecular gas phase \citep{Grenier05, Rollig06, Wolfire10, Glover12, Bolatto13, Accurso17a, Madden20}. Other emission lines, however, such as the far-infrared \ciiline\ line, are strikingly enhanced on global scales in dwarf galaxies \citep[e.g.,][]{Cormier15, Cigan16, Lebouteiller17, Jameson18, Cormier19}, making the \ciiline\ line a potential tool for tracing star formation activity \citep{Malhotra01, Papadopoulos07, Pineda14, DeLooze14, Olsen15, Herrera15, Herrera18b, Carniani18, Matthee19, Izumi21, Fujimoto21} and a tracer of the total \hmol\ in galaxies, near and far \citep{Poglitsch95, Wolfire10, Pineda13, Nordon16, Fahrion17, Accurso17b, Zanella18, Madden20, Schaerer20, Tacconi20}.

In this paper, we explore the use of different infrared (IR) tracers, in particular the \oiiiline\ line together with the midinfrared (MIR) continuum at 24\mic, as a new method to understand the evolution of the warm dust and gas phases in galaxies. This new tool can also be used to prepare future observations, both at high redshift and in local galaxies. For the rest of the paper we make use of the following shorthand notation: \oiii\ $\equiv$ \oiiiline; \cii\ $\equiv$ \ciiline; and \oilinelo\ $\equiv$ \oi. 

The most luminous farinfrared (FIR) line observed in low-metallicity star-forming galaxies is the \oiii\ line, even brighter than the classically used \cii\ line, which is the brightest FIR line in metal-rich galaxies \citep{Stacey91, Malhotra01, Luhman03, Brauher08, Smith17, Diaz-Santos17, Croxall17, Sutter19, Sutter21}. The energy needed to create \oplusplus\ (35 eV) makes it a direct tracer of the ionized gas around young stars and thus an important coolant of the ionized medium and a SFR tracer \citep[e.g.,][]{DeLooze14, Herrera15}. The predominance of the \oiii\ line over full dwarf galaxy scales \citep{Cormier15,Cormier19} is an indication of an extensive filling factor of ionized gas in low metallicity environments, and can explain the potential for large scale photodissociation of the CO molecule in dwarf galaxies. The \oiii\ line has been observed in surveys of local galaxies of a wide range of metallicities with \hers\ \citep[e.g., DGS; KINGFISH:][]{Kennicutt11, Madden13, Lamarche17, Pereira17, Fernandez17} and is becoming a line frequently observed in high-z galaxies \citep[e.g., ][]{Ferkinhoff10, Matthee17, Carniani17, Tamura19, Hashimoto19a, Hashimoto19c, Hashimoto20}.

Often associated with warm dust in the ionized phases in galaxies, the MIR continuum at 24\mic, has been mapped with \spitz/MIPS in many galaxies \citep[e.g.,][]{Draine07, Bendo07, Bendo08, Bendo12, Croxall12, Vutisalchavakul13, Kendall15, Boquien16} including full maps of the nearby Magellanic Clouds \citep[e.g.,][]{Meixner06,Gordon11}. Assumed to be a tracer of dust in \HII\ regions, it is often used as a SFR tracer \citep[see e.g., ][ and references therein]{Kennicutt12}, but it can also originate in dust heated by older stellar populations \citep[e.g.,][]{Leroy12}. 

Therefore, most of the \oiii\ and  24\mic\ emission is associated with recent star formation in galaxies, suggesting that there should be an observational relationship between them. This paper investigates the link between \oiii\ and  24\mic\ for the first time and explores the origins of the contributions as well as the deviations of the relationship between these tracers. We investigate the potential to predict the \oiii\ line from the 24\mic\, band, with spatially resolved and integrated data (also to be used to predict 24\mic\, from \oiii\ observations). We also investigate the accuracy of the prediction and what drives deviations from the global behavior.

In order to understand what those relations on global scales, it is important to obtain a precise understanding of the properties and relation to the local environment of gas and dust phases at a resolved scale, and the consequences on the observed tracers. Only in this way can one be confident in employing certain diagnostic tracers for unresolved high-z studies as well. The closest low metallicity galaxies of the Milky way are the Large and Small Magellanic Clouds (respectively LMC and SMC). Their proximity (50 kpc for the LMC, \citealp{Walker12}, 61 kpc for the SMC, \citealp{Hilditch05}) gives us access to details of the properties of their star-forming regions, and allows us to disentangle the different phases of the ISM under low metallicity environments (Z$_{\rm LMC}$=1/2 \zsol, \citealp{Pagel97,Pagel03}, Z$_{\rm SMC}$=1/5 \zsol, \citealp{Russel92,Cioni00}). Thus, the Magellanic Clouds are the ideal laboratories to investigate the relation between \oiii\ and  24\mic\ in detail.

This paper is structured as follows. In Section \ref{sect-data}, we present the \hers\ and \spitz\ data used in the study. A direct relation between \oiii\ and the MIPS 24\mic\, band is studied in Section \ref{sect-direct_relation}. In Section \ref{sect-models} we present the models prepared for our analysis, and the specific technique we use to disentangle the emission from \HII\, regions and PDRs. The dependency with different physical parameters is analyzed in Section \ref{sect-physics}, where we also present a new method developed for this study allowing investigation of the effects of various physical parameters, including the proportion of PDRs needed in the models to reproduce the observations. This method can be seen as a step toward understanding the 3D geometry of the ISM. The method and the analyses are discussed in Section \ref{sect-discussion}, and we conclude our work in Section \ref{sect-summary}.


\section{Data}\label{sect-data}

The regions studied here are driven by the available \hers/PACS spectroscopy \citep{Pilbratt10, Poglitsch10} from the DGS survey, where maps of some FIR fine structure lines were observed for numerous star-forming regions of the LMC and SMC (Table \ref{tab-PACS_regions};  \citealp{Cormier15}). To this we added \spitz\ 24\mic\ photometry \citep{Werner04, Houck04, Rieke04} from the SAGE project \citep{Meixner06} which mapped the LMC and SMC completely.

\begin{table}
\caption{Star-forming regions of the LMC and SMC for this study.}\label{tab-PACS_regions}
\begin{center}
\begin{threeparttable}
\begin{tabular}{cc}
\hline
Region & Center\\
\hline
30 Doradus & 5h38\arcmin 43\arcsec, -69$^{\circ}$05\arcmin 48\arcsec \\
LHA 120-N11B & 4h56\arcmin 54\arcsec, -66$^{\circ}$24\arcmin 24\arcsec \\
LHA 120-N11C & 4h57\arcmin 43\arcsec, -66$^{\circ}$27\arcmin 40\arcsec \\
LHA 120-N11I & 4h55\arcmin 44\arcsec, -66$^{\circ}$34\arcmin 24\arcsec \\
LHA 120-N44 & 5h22\arcmin 03\arcsec, -67$^{\circ}$57\arcmin 52\arcsec \\
LHA 120-N158 & 5h39\arcmin 14\arcsec, -69$^{\circ}$30\arcmin 14\arcsec \\
LHA 120-N159 & 5h39\arcmin 54\arcsec, -69$^{\circ}$45\arcmin 18\arcsec \\
LHA 120-N160 & 5h39\arcmin 42\arcsec, -69$^{\circ}$38\arcmin 52\arcsec \\
LHA 120-N180 & 5h48\arcmin 39\arcsec, -70$^{\circ}$05\arcmin 54\arcsec \\
LHA 115-N66 & 0h59\arcmin 09\arcsec, -72$^{\circ}$10\arcmin 46\arcsec \\
\hline
\end{tabular}
\begin{tablenotes}
\item The center positions for each region are based on the \cii\ \hers/PACS observations.
\end{tablenotes}
\end{threeparttable}
\end{center}
\end{table}

\subsection{{\it Spitzer} data}\label{sect-data_spitzer}

We used the \spitz\ IRAC and MIPS maps available on the IRSA Infrared Science Archive\footnote{https://irsa.ipac.caltech.edu/data/SPITZER/SAGE/} and selected subregions covering our \hers\ spectroscopy maps. Two regions present some saturation in the MIPS 24\mic\ band: 30 Doradus and N160. For 30 Doradus, the MIPS 24\mic\, data were replaced with \spitz/IRS observations, integrated over the transmission profile of the MIPS band \citep{Chevance16}. As N160 was not observed with the IRS, the part of the map affected by saturation was simply masked.

Visual inspection showed that the 24\mic\ maps are strongly affected by point source emission, often corresponding to stellar sources. As we are interested in the ISM properties, we removed the point sources from the MIPS 24\mic\, map. Positions of the point sources were first selected from dedicated catalogs \citep{Seale12, Gruendl09, Jones17}; among these, sources to be removed were selected as those also detected as point sources in the IRAC 8\mic\ band. Verification was made by checking the presence of point source emission at the selected positions in the IRAC maps. A 2D point spread function (PSF) was fitted on the point sources, and removed from the 24\mic\ maps. Then, the background was fitted, assumed to be flat on tiles with a size depending on the PSF and the intensity of the point-sources. The aim was to avoid degeneracy in the extraction of the sources. For two regions we did not make this correction. For N160 the saturated part of the map was already masked, and all the point sources lie within this masked area. The N66 region contains a large number of point-sources, that lead to numerous artifacts when removing them, making the map impossible to use. Thus, we decided to keep the MIPS 24\mic\ map of N66 without correction for the point-sources. Keeping the 24\mic\ emission corresponding to point sources in this region does not significantly affect the global results in our study, but increases the uncertainty on the examined relationships.

The limiting spatial resolution for the correlations investigated in this study is due to the MIPS 70\mic\, observations, which have a beam full width at half maximum (FWHM) of 18\arcsec. Thus, the maps of all the tracers were convolved to 18\arcsec\, with a Gaussian kernel and resampled to a grid with pixels of 12\arcsec\, width in order to avoid oversampling of the beam.

\subsection{{\it Herschel} data}\label{sect-data_herschel}

The {\it Herschel}/PACS spectroscopic maps of LMC and SMC regions were initially presented in \cite{Cormier15}, along with other spectroscopic observations of the DGS sample. The processing of the data is fully described in \cite{Cormier15}. In this study, data were reprocessed with the latest version of HIPE \citep[v.15, ][]{Ott10} and PACSman \citep[v3.63, ][]{Lebouteiller12b}. The maps were also convolved with a Gaussian kernel to match the MIPS 70\mic\ resolution and similarly resampled to match the 24\mic\ maps.

For further comparison of our work with other dwarf galaxies, we used DGS \citep{Madden13}, which targets 48 dwarf galaxies of the local Universe, covering a large range of conditions, including metallicity and SFR. Indeed, the metallicity of the DGS sample ranges between 1/50 \zsol\ and 1/2 \zsol, with the LMC and SMC being at values of 1/2 \zsol\ and 1/5 \zsol, respectively. The SFR of the DGS varies widely, from 10$^{-3}$ M$_{\odot}$/yr to 100 M$_{\odot}$/yr, with values of $\sim$0.2 M$_{\odot}$/yr and $\sim$0.04 M$_{\odot}$/yr for the LMC and SMC respectively. The distances range between 50kpc and 60kpc for the LMC and SMC, repectively to nearly 200 Mpc for the furthest target of the DGS sample \citep[see][and references therein for further details]{Meixner13, Madden13}.

\section{Relation between \oiii\ and MIPS 24\mic}\label{sect-direct_relation}

We first investigated if there is a relation between surface flux density of \oiii\ and 24\mic. In general, this comparison could be done for any line unambiguously directly tracing \HII\ regions. We used \oiii\ here since this is a line frequently observed at both high and low redshift, especially with \hers\ and ALMA surveys. We also looked into the accuracy of the relationship between these tracers and, in cases where there may be deviations from the global behavior, into what drives the differences.

\subsection{Spatially resolved relation}\label{sect-direct_relation_resolved}

\begin{figure}
\begin{center}
\includegraphics[scale=0.5]{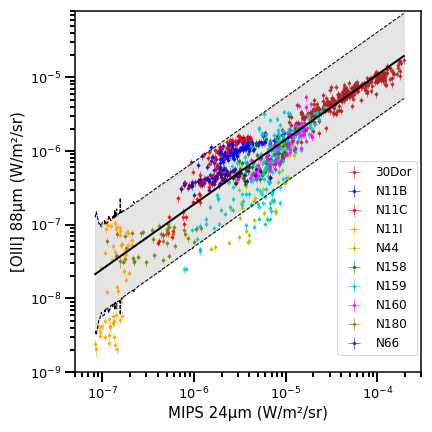}
\caption{Relation between \oiii\ and 24\mic\, in all of our SMC and LMC regions, only including pixels above 5$\sigma$ in the \oiii. The black solid line represents the fitted relation, linear in log-log space, and the dashed lines and gray filling represent the 95\% confidence interval of the fit. The colored symbols refer to the individual regions labeled in the legend to the right of the plot.}
\label{fig-OIIIbasic_correlation}
\end{center}
\end{figure}

We took advantage of the excellent spatial resolution of \hers/PACS and \spitz/MIPS together with the proximity of the Magellanic Clouds to investigate the relation between the \oiii\ and 24\mic\ at  $\sim$ 5\,pc scale. Figure \ref{fig-OIIIbasic_correlation} shows the correlation between the observed 24\mic\ continuum and the \oiii\ line at a spatial resolution of $12$\arcsec pixels in all of our sources for which the \oiii\ line was detected at more than 5$\sigma$. We see that the different sources together exhibit a wide range of values, spanning around 3 orders of magnitude in 24\mic\, and 4 orders of magnitude in \oiii. We found a clear linear relation between those two tracers in log-log space. This correlation is fitted with a weighted $\chi^{2}$ method, leading to the following relation:

\begin{equation}
\log(\Sigma_{\rm [OIII]\,88}) = 0.879 (\pm 0.012) \times \log(\Sigma_{\rm 24}) -1.446 (\pm 0.065).
\label{equ-basic_fit}
\end{equation}

\noindent where the surface brightness, $\Sigma$, is expressed in \unitsurf. We see that the spread varies significantly with the flux. It is tight (around 0.5 dex) for the brightest regions, especially 30 Doradus, N160, and the brightest regions in N159, but it can reach almost 2 dex deviations in other regions. This spread is partly due to differences between the local physical properties of the regions (e.g., between N44 and N11C), and intrinsic variations within each region (e.g., N159).

\begin{figure}[h!]
\begin{center}
\includegraphics[scale=0.5]{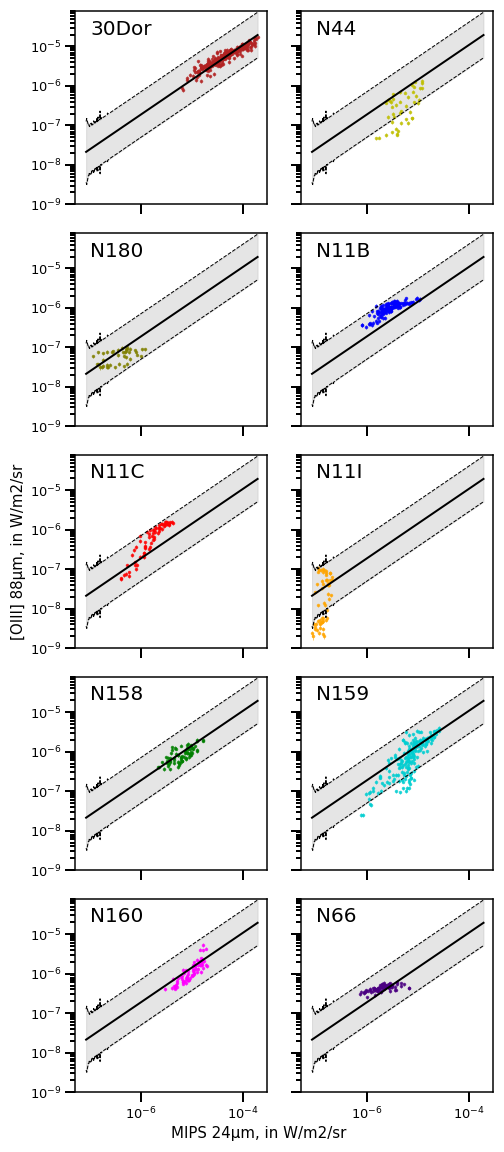}
\caption{Relation between surface brightness of \oiii\ and 24\mic\, for all of the pixels of our individual star-forming regions detected at more than 5$\sigma$. The black solid line represent the fitted linear relation in log-log space for the total sample (Figure \ref{fig-OIIIbasic_correlation}), and the dashed lines and gray filling represents the 95\% confidence interval associated with the total sample.}
\label{fig-OIIIbasic_correlation_mozaic}
\end{center}
\end{figure}

\begin{figure}
\begin{center}
\includegraphics[scale=0.4]{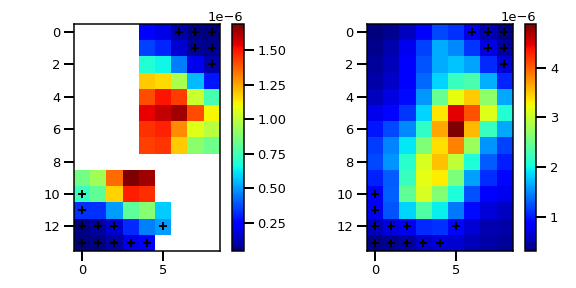}
\caption{Spatial distribution of the low emission pixels in N11C. Left panel: \oiii\ map for N11C; right panel: MIPS 24\mic\, map for N11C. Both maps are convolved to 12\arcsec\ and resampled to 12\arcsec\ pixels. The black crosses correspond to pixels that have low surface brightness (on the left of the bulk of the regions in Figures \ref{fig-OIIIbasic_correlation} and \ref{fig-OIIIbasic_correlation_mozaic}, corresponding to values of 24\mic\ lower than 10$^{6}$ W/m$^2$/sr) showing a different behavior from that of the global relation.}
\label{fig-N11C_visual}
\end{center}
\end{figure}

If we consider each region separately (Figure \ref{fig-OIIIbasic_correlation_mozaic}), we see that the slopes of most of the individual regions are close to the global slope which seems to be driven by 30Dor. All except a few positions in N44, N11I and N159 fall within the 95\% confidence level found from the global relation, with the deviant pixels often being low brightness pixels. In N11C (Figure \ref{fig-OIIIbasic_correlation_mozaic}), for example, the slope begins to steepen when the 24\mic\, surface brightness is less than $\sim$10$^{-6}$ \unitsurf. These particular low-surface brightness pixels lie on the outskirts of the star-forming peak in N11C (see Figure \ref{fig-N11C_visual}). Thus, we see that the \oiii-24um relation depends on the surface brightness. Moreover, we see that the regions are distributed along the relation, some of them showing strong emission of both \oiii\ and 24\mic\ (e.g., 30 Doradus, N159), and some showing weak emission of both tracers (e.g., N11I, N180). This variation could be linked to the age of the ionizing source. A younger cluster will produce more ionizing photons, which increases the emission of the \oiii\ line, and it will be surrounded by a more compact shell of dust, resulting in an increase in the 24\mic\ emission. The two tracers are probably not affected similarly, thus the relation may be affected by the age of the star-formation burst. On the other hand, 24\mic\ emission can be powered by an older population of stars, especially in regions that have sustained multiple episodes of star formation \citep{Leroy12}. Thus, part of the spread of the data around the relation could be due to 24\mic\ emission arising from older stars, especially in more quiescent regions such as N11I or N180.

N66 is the only region in the SMC in this study. Both the \oiii\ and 24\mic\, may be affected by the lower metallicity, which can globally decrease the amount of dust and \oiii\ in the gas phase simply owing to the lower metal abundance. On the other hand, as the dust abundance is diminished, the ISM may be subjected to an overall harder radiation field, which may increase the 24\mic\, emission by shifting the peak of the SED to shorter wavelengths \citep[see e.g.,][and references therein]{Galliano18b}. Despite those effects, the \oiii\ - 24\mic\ relation in N66 does not differ much from that of the LMC regions. The slope of the region appears flatter (Figure \ref{fig-OIIIbasic_correlation_mozaic}) than that of the global relation pointing to the possibility that \oiii\  and 24\mic\, may not be as closely linked in the SMC as in the LMC regions. For example, the \oiii\ in N66 may originate from a more extended, diffuse component that is not completely associated with compact star-forming regions from which the 24\mic\, originates. More studies on other regions in the SMC would be necessary to conclude on this point.

We note that the fits to the \oiii\ and 24\mic\, correlations were made using convolved and resampled maps, and the MIPS 24\mic\, maps were treated to remove contamination from point sources (cf Section \ref{sect-data_spitzer}). Any of these steps could possibly influence the fitting results. In Appendix \ref{appdx-test_fit} we detail the effects of different fitting methods and different data treatment and point-source removal methods: a) effect of different fitting methods: comparison of a simple, nonweighted $\chi^2$, and a Monte Carlo simulation to determine uncertainties and the weighted $\chi^2$ fit; b) effect of resolution and sampling of the maps on the fits; c) tests of the method for point source subtraction and corrections. We conclude from the different tests that the fit remains robust in these different cases.

The spatially resolved data in Figure \ref{fig-OIIIbasic_correlation} show a clear, linear relation between \oiii\ and 24\mic\, in log-log space, which can be used, for example, to predict \oiii\ line emission from maps of the 24\mic\, band, or inversely. The predictive power of this relation and the effect of the spread around the relation are studied in detail in Section \ref{sect-predictive_use}. However, the LMC and SMC are the closest neighbors of the Milky Way, and most of the local Universe galaxies will not provide such spatially detailed observations. We thus explored if the relation also holds for unresolved regions, to see if the correlation can be applied in a more general case and to more distant galaxies.

\subsection{Integrated relation and comparison with the DGS}

\begin{figure}[h!]
\begin{center}
\includegraphics[scale=0.5]{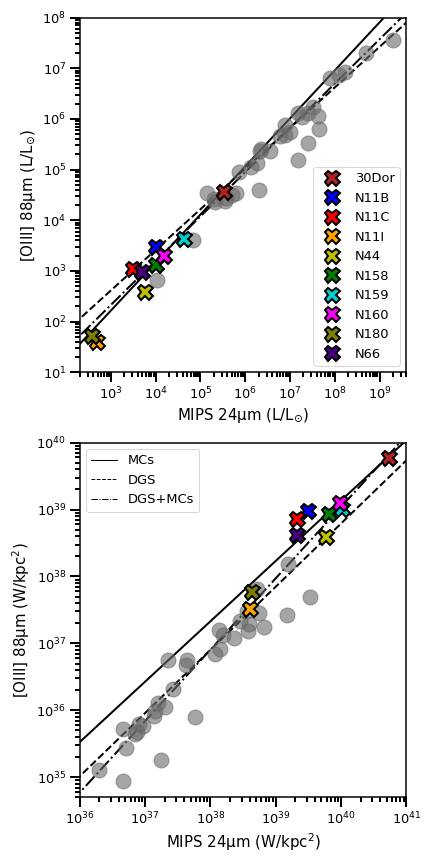}
\caption{Upper panel: Relation between the integrated \oiii\ and 24\mic\, luminosity, in units of solar luminosity (L$_{\odot}$). Lower panel: Integrated luminosity per surface area of \oiii\ and 24\mic\, (W${kpc}^{-2}$). The DGS sample corresponds to the gray circles, and the integrated regions from the Magellanic Cloud sample are coded with colors, as indicated in the inserted legend in the upper panel. The fit on the Magellanic Clouds alone is indicated as MCs, the fit on DGS sample is indicated as DGS, and the fit on the whole data is indicated as DGS+MCs.}
\label{fig-OIII_vs_24_unresolved_relation_DGS}
\end{center}
\end{figure}

In order to mimic unresolved observations, we used the spatial maps of the regions at their original resolution and sampling, without correction from point sources. The surface brightness of the complete regions were then calculated over the zones where the observations of the two tracers, \oiii\ and 24\mic, overlapped. To take into account the different distances of the Magellanic Clouds and the DGS galaxies, we converted the flux values into luminosities, normalized to solar luminosity, \lsol. The uncertainties of the integrated surface brightness values were calculated via a quadratic sum, and converted in the same way as the integrated luminosity values. We see in Figure \ref{fig-OIII_vs_24_unresolved_relation_DGS} that the integrated luminosities of the different regions of the Magellanic Clouds also show a linear trend in log-log space. Using a weighted $\chi^2$ method on those star-forming regions, we obtained the relation of Equation \ref{equ-lux_fit} for the integrated Magellanic Cloud regions:

\begin{equation}
\log(L_{\rm [OIII]\,88}) =  0.948 (\pm 0.086) \times \log(L_{\rm 24}) -0.622 (\pm 0.360),
\label{equ-lux_fit}
\end{equation}

\noindent with $L_{\rm [OIII]\,88}$ being the luminosity of \oiii\ and $L_{\rm 24}$ being the luminosity of the 24\mic\, continuum, both expressed in \lsol. We see that the slope obtained for the integrated luminosities of the Magellanic Clouds is consistent with that obtained for surface brightness values (Equation \ref{equ-basic_fit}) within the uncertainties. We also investigate the relation with the surface luminosity.

\begin{table}
\begin{center}
\caption{Fit parameters of the relation between \oiii\ and 24\mic\ from Figure \ref{fig-OIII_vs_24_unresolved_relation_DGS} for the integrated Magellanic Clouds regions (MCs), the DGS sample alone, and the DGS sample plus the integrated Magellanic Clouds regions (DGS + MCs), with 1$\sigma$ uncertainty on the slope and intercept.}
\begin{tabular}{ccc}
\hline
Sample & Slope & Intercept\\
\hline
\hline
$L$/L$_{\odot}$ & &\\
\hline
MCs & 0.948 $\pm$ 0.086 & -0.622 $\pm$ 0.360\\
DGS & 0.801 $\pm$ 0.036 & 0.215 $\pm$ 0.242\\
DGS + MCs & 0.873 $\pm$ 0.021 & - 0.288 $\pm$ 0.128\\
\hline
\hline
$W$/kpc$^{2}$ & &\\
\hline
MCs & 0.903 $\pm$ 0.115 & 3.024 $\pm$ 4.565\\
DGS & 0.945 $\pm$ 0.076 & 0.997 $\pm$ 2.924\\
DGS + MCs & 1.068 $\pm$ 0.052 & -3.667 $\pm$ 2.041\\
\hline
\end{tabular}
\label{tab-DGS_fullsample_fit}
\end{center}
\end{table} 

In Figure \ref{fig-OIII_vs_24_unresolved_relation_DGS} we also compare the \oiii\ and 24\mic\, luminosities of the integrated Magellanic Cloud regions with those of the integrated DGS targets. They are normalized by the solar luminosity (upper panel) or by the surface (lower panel). We examined three different fits, one on our Magellanic Cloud integrated sample, one on the DGS sample and one on all the data (Table \ref{tab-DGS_fullsample_fit}). We see that the different fits give close values, agreeing with each other considering the uncertainties. However, in the upper panel, some of the DGS targets lie significantly below the fit on the Magellanic Clouds regions, and the fit on the DGS sample has a somewhat shallower slope. As discussed in Section \ref{sect-discussion}, this may indicate a small enhancement of 24\mic\, emission compared to \oiii\ on the scale of total galaxies. This effect is more evident in the lower panel, where the luminosity per surface area is shown. Indeed, \oiii\ arises from energetic ionized media heated by young O and B stars, while 24\mic\, emission can be powered by older stellar populations as well, more prominently when integrating over full galaxy scales. \cite{Leroy12} studied the proportion of 24\mic\ not associated with recent star formation in a sample of 30 local disk galaxies. They found a median value of 19\% of the total 24\mic\ emission that is linked to older stellar populations, and possibly up to 40\% in some parts of the galaxies. The DGS sample we use here is a collection of star-forming dwarf galaxies, known to have a bursty star formation history, in contrast to \cite{Leroy12} sample of disk galaxies for which a more continuous star formation is invoked. The effect of prominent recent SF and lower dust abundance in this dwarf galaxy sample \citep[e.g.,][]{Remy14, Galliano18b} leads to a more porous ISM for hard UV photons globally \citep[see e.g., ][]{Galliano03, Galliano05, Madden06, Cormier15, Cormier19, Madden20, Ramambason22}. Thus, the proportion of 24\mic\ emission not associated with recent star formation is probably lower than that found in disk galaxies.
Some of 24\mic\, not arising from the neighborhood of young star-forming regions may thus be linked to the presence of non star-forming regions within the beam. However, this difference is small, suggesting that most of the 24\mic\, emission is still associated with star-forming regions, even when considering the full emission of a galaxy \citep[e.g.,][]{Contursi02, Calzetti05, Alonso06, Perez06, Prescott07}. Our small maps of Magellanic Cloud regions are centered on the brightest zones of star formation, largely filled by ionized gas \citep[see e.g., ][]{Kurt98, Naze01, Barba03, Lebouteiller12b, Chevance16, Gordon17, McLeod19}. If we could observe more extended regions within the Magellanic Clouds, similar to full galaxy scales, we might collect more PDR gas within our beam and possibly some 24\mic\, emission from PDRs as well as older stellar populations, which would then further enhance the 24\mic\ emission compared to \oiii\ emission. Despite this caveat, this relation holds over more than three orders of magnitude for both \oiii\ and 24\mic, and is still valid for numerous galaxies at different distances from the Milky Way, giving our calibration a potentially powerful predictive potential, that we demonstrate in the following section.

\subsection{Predictive power of the \oiii\ - 24\mic\ calibration}\label{sect-predictive_use}

We used our two calibrations from Equations \ref{equ-basic_fit} and \ref{equ-lux_fit} to explore the predictive potential of the observed correlation between 24\mic\ and \oiii, and we compared them to the resolved and unresolved observations. We calculated the uncertainties of the predicted values with a Monte Carlo simulation, using the uncertainties on the fitted parameters. Those uncertainties are compared to the difference between the observed and predicted values of \oiii. 

\begin{figure}[h!]
\begin{center}
\includegraphics[scale=0.5]{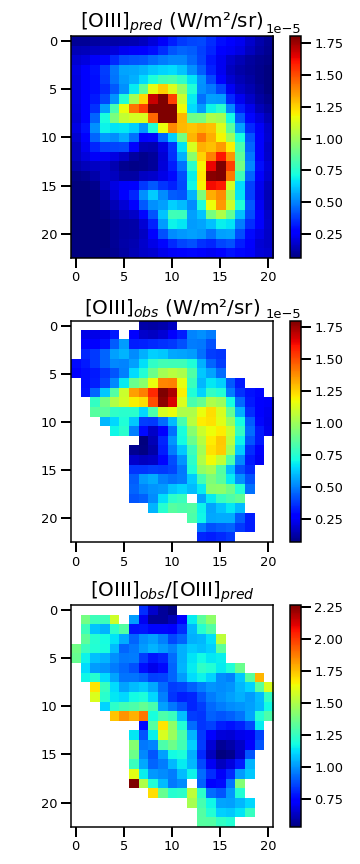}
\caption{The predicted 30 Dor map of \oiii\ (top panel) given the observed 24\mic\, and Equation \ref{equ-basic_fit}. For comparison, the actual observed map of \oiii\ (middle panel), and the ratio of observed over predicted emission (lower panel). The two maps of \oiii\ are represented on the same scale and are displayed in \unitsurf.}
\label{fig-OIIIpredicted_30Dor}
\end{center}
\end{figure}

\begin{figure}[h!]
\begin{center}
\includegraphics[scale=0.5]{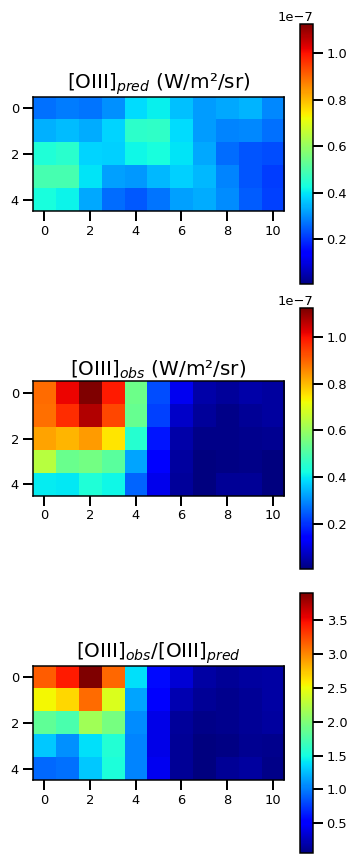}
\caption{Similar to Figure \ref{fig-OIIIpredicted_30Dor}, but for N11I.}
\label{fig-OIIIpredicted_N11I}
\end{center}
\end{figure}

We first compared the \oiii\ predictions we derive from the observed 24\mic\ using surface brightness, Equation \ref{equ-basic_fit}, to the observed \oiii\ maps of the Magellanic Cloud regions. As examples, we show in Figures \ref{fig-OIIIpredicted_30Dor} and \ref{fig-OIIIpredicted_N11I} the comparison for the 30Dor and  N11I regions, respectively. The predicted and observed maps for the other regions are shown in the Appendix \ref{appdx-regions_plots}.

\begin{table}
\begin{center}
\caption{Summary of the proportion of our data for which the predicted and observed \oiii\ emission agree within a factor of 3, and the range of values of the ratio \oiii $_{\rm obs}$/\oiii $_{\rm pred}$.}
\begin{tabular}{cccc}
\hline
 & Percentage of & Range of values of \\
 & points agreeing & \oiii $_{\rm obs}$/\oiii $_{\rm pred}$ \\
 & within a factor of $3$ & \\
 & (\%) & \\
\hline
All regions & 90 & 0.06 - 4.15\\
30 Dor & 100 & 0.53 - 2.26 \\
N11B & 95 & 1.02 - 4.07 \\
N11C & 81 & 0.57 - 4.15 \\
N11I & 47 & 0.06 - 3.91 \\
N44 & 53 & 0.10 - 1.21 \\
N158 & 100 & 0.52 - 2.05 \\
N159 & 82 & 0.11 - 3.40 \\
N160 & 100 & 0.51 - 2.45 \\
N180 & 98 & 0.30 - 2.06 \\
N66 & 100 & 0.42 - 2.27 \\
\hline
\end{tabular}
\label{tab-pred_consistency}
\end{center}
\end{table}

The predicted and observed \oiii\ emission agree within a factor of $3$ for 90\% of the full data set, with a global discrepancy not higher than 1 dex, and for all of the regions but two, the predicted and observed \oiii\ agree within this factor of 3 for more than 80\% of the data set (see Table \ref{tab-pred_consistency}). We also point out that the intensity of \oiii\ does not seem to strongly affect the goodness of the prediction down to $3 \times 10^{-8} $W/m$^2$/sr. Only N11I, which is the faintest region in both 24\mic\ and \oiii, exhibits \oiii\ emission under this value. This is also one of the two regions for which only $\sim$50\% of the data shows agreement between observations and predictions within a factor of 3, and shows a large discrepancy between the observed spatial distributions of \oiii\ and 24\mic (see Figure \ref{fig-OIIIpredicted_N11I}). This region seems to be more quiescent than the rest of the sample, and \oiii\ and 24\mic\ emission do not seem to be tightly linked, leading to different spatial distributions.

Other regions with significant differences between predictions and observations are N11B and N11C (see Figure \ref{fig-OIII_predovobs}, and Figures \ref{fig-app_N11B} and \ref{fig-app_N11C} in Appendix \ref{appdx-regions_plots}). These regions lie slightly above the global relation in Figure \ref{fig-OIIIbasic_correlation_mozaic}, although they have similar slopes. Table \ref{tab-pred_consistency} also shows that those regions exhibit the highest values of \oiii $_{obs}$/\oiii $_{pred}$. Likewise, we see from Figure \ref{fig-OIIIbasic_correlation_mozaic} that N44 and N159 have numerous pixels with observed 24\mic\ and \oiii\ values lying below the global relation. In particular, only 53\% of the N44 observed emission agrees within a factor of 3 with the predictions. The exploration of the physical conditions in the different regions in Section \ref{sect-physics} shows that the physical parameters are significantly different in this region compared to the rest of the sample. The offsets of some regions from the observed global correlation (\eg, N11B) leads to under or overestimation of the \oiii\ emission by a factor of a few over most maapped areas. It can be due to differences in physical conditions within the individual regions, which is investigated in Section \ref{sect-individual_regions}. Nevertheless, our correlation allows us to make predictions of \oiii\ from the observed 24\mic\, which are accurate up to a factor 3.

\begin{figure}[h!]
\begin{center}
\includegraphics[scale=0.5]{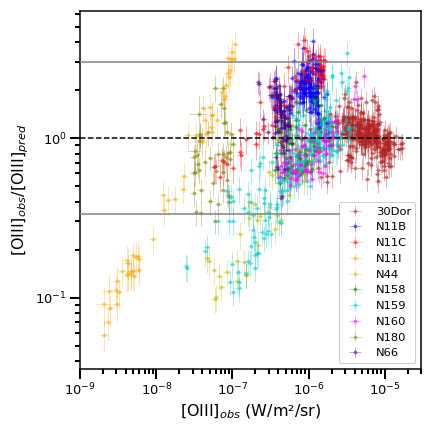}
\caption{Spatially resolved ratios of observed (\specie{O}{iii}$_{\rm obs}$) over predicted (\specie{O}{iii}$_{\rm pred}$) emission of \oiii. The value of 1, corresponding to good agreement of the predictions and the observations, is indicated with a horizontal dashed line. The horizontal gray lines indicate differences of a factor 3 from agreement.}
\label{fig-OIII_predovobs}
\end{center}
\end{figure}

\begin{figure}[h!]
\begin{center}
\includegraphics[scale=0.5]{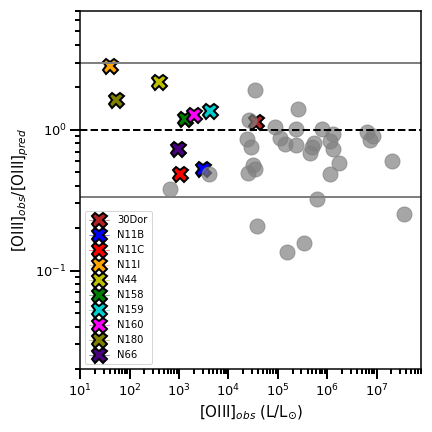}
\caption{Predicted \oiii\ luminosity, from the observed 24\mic\, for integrated luminosities of the Magellanic Clouds sample and DGS galaxies compared to the observed luminosity values. The horizontal dashed line indicates  when predicted and observed values are equal.The horizontal gray lines indicate differences of a factor 3 from agreement}
\label{fig-OIII_pred_DGS}
\end{center}
\end{figure}

We also investigated the predictions for the integrated values of our sample and the DGS values, based on the luminosities. Here again, predicted and observed \oiii\ luminosities agree within a factor of 3 for the Magellanic Cloud sample and for all the DGS targets (Figure \ref{fig-OIII_pred_DGS}) except four. The discrepancies between predictions and observations are similar to those found for the resolved data, with a factor of a few for most of the regions, and a couple of extreme DGS galaxies for which the disagreement can be up to almost 1 dex. To summarize, the calibrations we derived here allowed to predict \oiii\ within a factor 3 for most of the data, in both surface brightness and luminosity, for resolved and unresolved data. They also hold for galaxies with different physical conditions and distances from the Milky Way see Section \ref{sect-data_herschel} for more details. Thus, the calibrations we derive here may be useful to prepare observations at both low and high redshift, from the scale of spatially resolved, individual star-forming regions to integrated galaxies. We point out that this calibration can be used in both ways, to also predict 24\mic\ emission, adding potentially useful continuum information for high-redshift galaxies where the \oiii\ is detected. 

However, we also showed that there is a nonnegligible spread in some cases, around the correlations derived from Figures \ref{fig-OIIIbasic_correlation} and \ref{fig-OIII_vs_24_unresolved_relation_DGS}, that can be due to differences in the overall physical conditions between the individual regions of our Magellanic Clouds sample, as well as locally within an individual region. In order to understand the spread around this relation, we built models to study the effect of some physical parameters. 

\section{Modeling strategy}\label{sect-models}

We used the spectral synthesis code, Cloudy (version 17.01, \citealp{Ferland17}) to interpret the observations. Cloudy is a 1D spherical photodissociation and photoionization code that selfconsistently solves the thermal, chemical, and radiation equilibrium of a multiphase medium exposed to a radiation source and produces a complete spectral energy distribution of the gas and dust. Input parameters include the spectrum of the central source of the radiation field and the total initial hydrogen density at the illuminated face of the cloud, \n. We chose the stellar population model PopStar \citep{Molla09}, which simulates the SED of a full stellar population based on a Chabrier initial mass function \citep{Chabrier03} at different ages, for a single burst of star formation. We did not test the assumption of a continuous star formation history because our observations are centered on small regions presenting recent star formation, thus it is likely that the ionizing clusters in our maps have similar ages. This is also consistent with previous studies on the stelar content of our regions \citep[e.g., ][]{Bica96, Heydari00, Caulet08, Carlson12, Ochsendorf17}. In our models, we supposed that the ISM is in pressure equilibrium, thus we set constant total pressure through the cloud. We also ran a set of models with density kept constant through the cloud, but it was not possible to reproduce all the observed tracers. We also noticed that characteristic tracers from PDRs were better reproduced by models with a higher density at the inner radius than those reproducing the characteristic tracers of \HII\ regions, which is consistent with constant pressure models, and is consistent with previous works comparing the different assumptions \citep[e.g.,][]{Tielens85,Hollenbach99,Bron18,Cormier19}.

We then varied two parameters: \n, the density at the inner radius, and the ionization parameter ($U$), which describes the effect of the ionizing source on the gas and dust. $U$ is the dimensionless parameter relating the number of hydrogen-ionizing photons emitted by the central source, $Q({\rm H^0})$, impacting the inner surface of the cloud of radius, $r$, the speed of light, $c$ and \n:

\begin{center}
\begin{equation}
U = \dfrac{Q({\rm H}^0)}{4\pi \, r^2 \, n_{0}\, c}.\label{equ-Udef}
\end{equation}
\end{center}

\n\ is varied in log(cm$^{-3}$) from $-1$ to $3$ with steps of $0.5$, and $U$ is varied in log space from $-4$ to $-1$ with steps of $1$. The age of the simulated cluster is varied between $1$ and $10$ Myr, with steps of $0.2$ in log space. The default dust abundance and polycyclic aromatic hydrocarbons (PAH) properties in Cloudy are those from the Milky Way. We applied a multiplicative factor to the abundances to match the SMC and LMC metallicities. In order to completely model the PDR up to the molecular region, the models were stopped at an \Av\ of 2.5, which ensures that the formation of the \hmol\ molecules in all our models is reached, which is considered as the beginning of the molecular region. We initially computed models for two metallicities, 1/2 Z$_{\odot}$ and 1/5 Z$_{\odot}$, corresponding to the conditions in the LMC and SMC respectively. However, the difference between the two sets of models in the outputs used in our work were not significant. Thus, we kept one set of models with the LMC metallicity, as most of our regions are from the LMC. Table \ref{tab-models_charac} summarizes the range of the parameters and the steps of each parameter used to create the grids. 

\begin{table}
\label{tab-models_charac}
\caption{Main parameters and their range of values used to compute Cloudy model grids for this study.}
\begin{tabular}{ll}
\hline
Parameters & \\
\hline
Metallicity & $1/2 Z_{\odot}$\\
$\log$ of age (yrs) & $6-7$, steps $0.2$ \\
$\log$(\n) (cm$^{-3}$) & $-1 - 3$, steps $0.5$\\ 
$\log$($U$)$^a$ & $-4 - -1$, steps $1$\\
Stopping criterion & $A_{\rm V}=2.5$\\
\hline
\multicolumn{2}{l}{(a) The ionization parameter, U, is defined in Equation \ref{equ-Udef}} \\
\end{tabular}
\end{table}

\begin{figure}[h!]
\begin{center}
\includegraphics[scale=0.37]{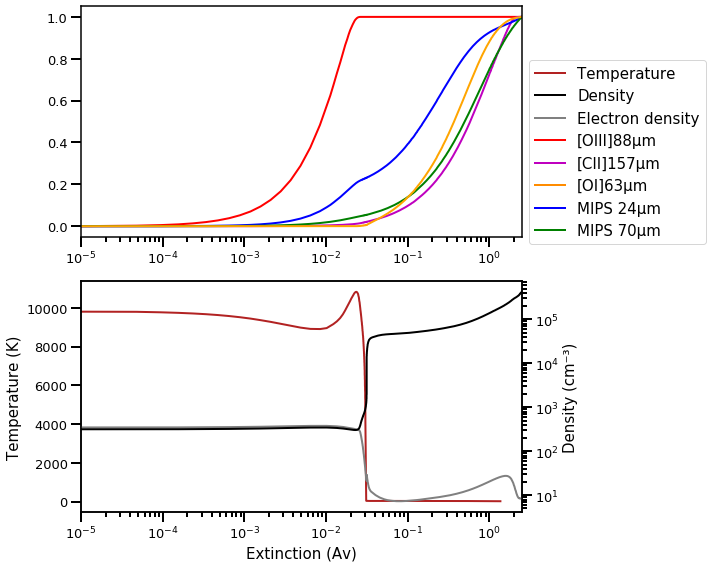}
\caption{Example from our Cloudy models. The cumulative emission of different bands and emission lines are presented (upper panel), together with some physical parameters (density, electronic density and temperature, lower panel) as a function of the depth in the cloud, measured with the parameter, A$_{\rm V}$. The emission of all of the tracers were normalized to a maximum of 1 to allow a direct comparison. The age of the central ionizing source is 4 Myr, and the initial values of physical parameters are $\log$(\n)$= 2.5$ and $\log(U) = -3$.}
\label{fig-model_example}
\end{center}
\end{figure}

Figure \ref{fig-model_example} is an example of the models we created with Cloudy, showing the evolution of different physical parameters and the cumulative emission of the tracers we use in our following study with the depth of the cloud (presented with A$_{\rm V}$ values). The limit between the \HII\ region and the neutral gas is fixed at the point where the electronic density drops below half its maximum value (A$_{\rm V} \sim\, 3\,\times\, 10^{-2}$ in our example). The density is low in the \HII\ region, and increases in the PDR (up to a value of $\sim  \log$(\n)$= 5.6$ in our example, to be compared to the initial value at the illuminated edge of the cloud of $\log(n)= 2.5$).

Concerning the lines and bands studied here, two immediate features are noteworthy. Firstly, we noticed that all of the \cii\ emission arises from the PDR in our models, not the \HII\ region. This is consistent with the global studies of low-metallicity dwarf galaxies where most of the \cii\ arises from PDRs \citep[e.g., ][]{Cormier19}. Secondly, we found in our models that a significant fraction of the 24\mic\, continuum emission arises from PDRs, while it is often implicitly assumed to trace warm dust in \HII\ regions \citep[see e.g., ][for this assumption]{Draine07, Croxall12}. These two points will be important for the comparison with data in Section \ref{sect-physics}.The ISM observed within our beam is composed of a mixture of different phases with different physical conditions (diffuse ionized regions, PDRs, denser \HII\ regions, etc.), and can not always be modeled by a single Cloudy model (i.e., 1D model simultaneously solving one \HII\, region plus one PDR). 

We saved the cumulative emission as a function of the depth inside the cloud to examine the variation of some tracers with the cloud depth. Then, we separated the emission into two parts: that arising from the \HII\, region, until the ionization front, and that arising from the PDR\footnote{Here we refer to the PDRs as the H{~\sc i} region including the PDR}, beyond the ionization front and up to A$_{\rm V}$=2.5. We considered the total emission from the \HII\, region as a pure \HII\, region model and the total emission from the PDR as a pure PDR model. We then mixed the two components to mimic a physical mixture of ionized region and PDR along the line of sight for each pixel in the map, using equation \ref{equ-general_intensity}:

\begin{equation}
\left\{
\begin{array}{ccl}
I_{\rm mixed\, model} & = & c_{\rm H\, II}\,I_{\rm H\, II} + c_{\rm PDR}\,I_{\rm PDR}\\
 & = & c_{\rm H\, II}\,I_{\rm H\, II+PDR} + (c_{\rm PDR}-c_{\rm H\, II})\,I_{\rm PDR}\\
c_{\rm H\, II}+c_{\rm PDR} & = & 1 ,
\end{array}
\right.
\label{equ-general_intensity}
\end{equation}

\noindent where I$_{\rm H\, II}$ is the intensity of the tracer in the \HII\, region only in W\,m$^{-2}{\rm sr}^{-1}$; I$_{\rm PDR}$ is the intensity of the tracer in the PDR only and I$_{\rm mixed\, model}$ is the total intensity of a tracer for the mixed model along the line of sight. \chii\ and \cpdr\ are the scaling factors of the \HII\, region and PDR respectively. As we keep their sum equal to 1 in our strategy, we solve for \cpdr\ in the following work. The multiplicative factor \chii\ for the full model (I$_{\rm HII+PDR}$) is similar to a luminosity scaling, but we prefered to use ratios between lines and continuum bands.

With this method, we illustrated different physical configurations for the distribution of \HII\, regions and PDRs by changing \cpdr, the proportion of PDR (and by construct, \HII\, regions) in Equation \ref{equ-general_intensity}. We note that this mixture is made only from models having the same initial \n\ and U. We used this simple approach to avoid degeneracy between the parameters, as a first step. A more complex use of our model will be left for future research. 

\begin{figure}[h!]
\begin{center}
\includegraphics[scale=0.37]{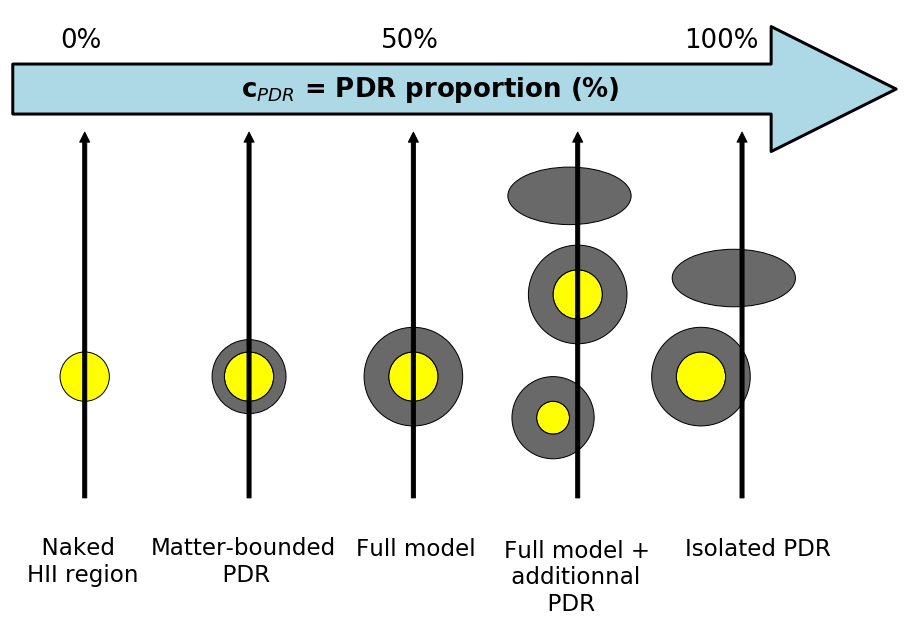}
\caption{Sketch of different PDR proportions (\cpdr) in our mixed models. The yellow regions represent \HII\, regions, and PDRs are represented in gray. The blue arrow indicates the direction of increasing \cpdr. The black arrows show the lines of sight through the represented clouds (each line of sight representing a given pixel in our maps). As per Equation \ref{equ-general_intensity}, the 0\% \cpdr\ case corresponds to no PDR in the model, hence fully ionized gas; 50\% \cpdr\ corresponds to a single Cloudy model, with a connected \HII\, region and PDR (Equation \ref{equ-PDR50}); and 100\% \cpdr\ represents a full PDR model, without an \HII\, region contribution. We note that 50\% of PDR does not correspond to a medium with half of the volume being filled by PDR, but corresponds to a mix that can be modeled by a Cloudy model with a full PDR layer.}
\label{fig-PDRproportion_sketch}
\end{center}
\end{figure}

Some of the cases that can be described by Equation \ref{equ-general_intensity} are illustrated in Figure \ref{fig-PDRproportion_sketch}. For example, when there is 0\% of PDR, the model corrspond to a naked \HII\, region. Similarly, a proportion of 100\% PDR corresponds to a model of an isolated PDR. We underline that in reality, the model of an isolated PDR component corresponds to an observation with only PDR emission along the line of sight, but the PDR can be a part of a larger structure that also contains \HII\, regions not seen along this line of sight. A proportion of 50\% \cpdr\ corresponds to a full Cloudy model (1D connected \HII\, region and PDR).

The extreme cases of 0\% and 100\% \cpdr\ are easy to represent, and all of the other cases are based on the interpretation of Equations \ref{equ-PDR10} to \ref{equ-PDR90} detailed in Appendix A. The \cpdr\ we used here does not correspond to the amount of PDRs in the medium, but to the amount of PDR emission from the model we used. A \cpdr\ greater than 50\% corresponds to a covering factor of the PDR on the \HII\, region greater than 1, with more PDRs along the line of sight than a simple model, or in other words there is more than one cloud along the line of sight, with potentially different intercepts between the cloud and the line of sight. Thus, it can be a way of interpreting the distribution of matter in the third dimension. 

In reality, the distribution of the matter in the ISM can be much more complex. Around a young star, neither the \HII\, region nor the PDR may fully cover the stellar source. The \HII\, region can be reduced to a clump of ionized gas near the star, and the covering factor of the PDR can be lower than unity, leading to a configuration with part of the cloud surrounding the star being matter-bounded \HII\ regions, where ionizing photons escape (no or little PDR beyond the ionized gas), and part of it being radiation bounded \HII\ regions where no photons with energies greater than 13.6 eV escape (with a full PDR beyond the ionized gas). This means that if along the line of sight there is a thick PDR covering only a part of the \HII\, region, we model it as a thinner PDR which is fully covering the \HII\, region. Higher spatial resolution observations, of course, would help to constrain a more detailed modeling scheme.

This method allowed us to constrain the input physical parameters of our Cloudy models: the age of the burst (Section \ref{sect-age}), \n\ and $U$ (Section \ref{sect-ISM_conditions}). Based on those parameters, it was possible to determine \cpdr\ for each pixel in the map. To constrain our mixed model, we used a ratio of observed PDR and \HII\, region tracers, such as \PDRHIIratio, to determine \cpdr\ needed in the model to reproduce the observed ratio, using Equation \ref{equ-PDRprop} as follows.

\begin{equation}
\begin{array}{ccc}
R_{\rm obs} & = & \frac{I_{\rm CII,\,obs}+I_{\rm OI,\,obs}}{I_{\rm OIII,\,obs}}\\
 & = & \frac{\left( c_{\rm HII}\, I_{\rm CII,\,HII}+c_{\rm PDR}\, I_{\rm CII,\,PDR} \right) + \left( c_{\rm HII}\, I_{\rm OI,\,HII}+c_{\rm PDR}\, I_{\rm OI,\,PDR}\right)}{c_{\rm HII}\, I_{\rm OIII,\,HII}+c_{\rm PDR}\, I_{\rm OIII,\,PDR}}\\
c_{PDR} & = & \left( 1-\frac{I_{\rm CII,\,PDR}+I_{\rm OI,\,PDR}-R_{\rm obs}\, I_{\rm OIII,\,PDR}}{I_{\rm CII,\,HII}+I_{\rm OI,\,HII}-R_{obs}\times I_{\rm OIII,\,HII}}\right)^{-1},
\end{array}
\label{equ-PDRprop}
\end{equation}

\noindent $I_{\rm CII,\,obs}$ the observed line intensity of \cii;  $I_{\rm OI,\,obs}$ is the observed line intensity of \oi\ and $I_{\rm OIII,\,obs}$ is the observed line intensity of \oiii, all along the same line of sight. We used Equation \ref{equ-general_intensity} in the ratio $R_{\rm obs}$ for each tracer in order to compare the models to the observations (Section \ref{sect-physics}), where $I_{\rm CII,\,HII}$ is the intensity of the \cii\ line in the \HII\, region model, with similar definitions for the other lines, and $I_{\rm CII,\,PDR}$ is the intensity of the \cii\ line in the PDR model, with similar definitions for the other lines. Equation \ref{equ-PDRprop} provides \cpdr\ which gives the final formula for the PDR proportion along each sight line. 

\section{Disentangling dust emission from the ionized phase and PDRs}\label{sect-physics}

We saw in Section \ref{sect-direct_relation} that the relation between \oiii\ and 24\mic\ emission shows a nonnegligible dispersion. Based on models presented in Section \ref{sect-models}, we investigated the possibility that part of the 24\mic\ may not arise from the same phase as \oiii. On the other hand, some regions are globally offset from the global relation, thus different physical conditions between the regions could also explain part of the spread around the relation. In order to examine those two potential effects and constrain our models, we used the \oiii\ line as a tracer of the \HII\ regions, and the \ciiline\ line and \oilinelo\ line as tracers of the PDR. In higher metallicity galaxies, the \cii\ line can arise from both \HII\ region and PDRs, but in low-metallicity dwarf galaxies, it mostly originates from PDRs (20 to 30\% at most coming from \HII\ regions, see e.g.,  \citealp{Cormier19, Tarantino21} and it is not greater than 5\% in the SMC according to \citealp{Jameson18}). We thus considered in this paper that the observed \cii\ emission comes from PDRs. The \oi\ line has a clearer origin, and is used to trace PDRs with high UV illumination and/or high density (see e.g., \citealp{Tielens85, Tielens05, Osterbrock06}).

We first investigated the physical conditions of the different regions, and especially the parameters we used to create our grid of models: age, \n\ and $U$. Then, we used Equation \ref{equ-PDRprop} to estimate in each pixel the contribution from PDRs and the ionized phase to the 24\mic\ emission. We used ratios of the tracers to be able to compare the different regions without requiring a calibration of the total luminosity, especially the ratio \OIIItfratio, to explore the contribution of each phase to the 24\mic\ emission, and the ratio \Rpdrhii\ = \PDRHIIratio, that we used as a proxy for the ratio of PDR over \HII\ region emission.

\subsection{Age of the burst}\label{sect-age}

\begin{figure}[h!]
\includegraphics[scale=0.4]{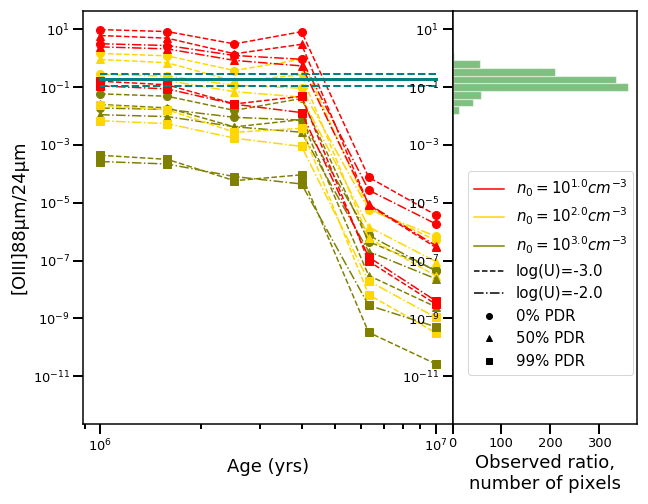}
\caption{\OIIItfratio\, for different models as a function of burst age. Each model is represented with a color for \n,  and a line style for U, and the lines are following the proportion of PDR emission, c$_{PDR}$, with the markers representing the \cpdr\ of 0\%, 50 \% and 99\%. The blue solid horizontal line corresponds to the median of the spatially resolved observations and the dashed blue lines to the median absolute deviation. The distribution of our observations is shown on the right as a histogram.}
\label{fig-OIIIov24_age}
\end{figure}

There are many studies on the star-forming regions on the Magellanic Clouds, and the age of recent star formation is frequently studied, mainly through the presence of O and B stars, or Young Stellar Objects \citep[YSOs; see e.g.,][]{Heydari86, Heydari02, Heydari10, Bolatto00, Naze01, Tsamis03, Gruendl09, Lebouteiller12b, Seale12, McLeod19, Okada19}. As we previously noted, the \oiii\ line is a characteristic tracer of the massive young stellar population, while the 24\mic\, emission can be produced by other stellar types. Thus, the ratio \OIIItfratio\, can be used to trace the overall age of the stellar population. We sampled ages between 1 and 10 Myrs, and compared the observations to the predicted line ratio \OIIItfratio\ for models that cover a range of \n, U and \cpdr\ (Figure \ref{fig-OIIIov24_age}).

From Figure \ref{fig-OIIIov24_age}, we see that the predicted \OIIItfratio\ ratio is roughly constant from 1 to $\sim$4 Myrs, and drops after this age, as the younger, more massive stars die off. Our observations show values that are consistent with the early age models. This is also consistent with several previous studies on these individual regions, proposing ages younger than 10 Myrs, often found to be ages of a few Myrs \citep[see e.g.,][]{Lucke70,Oey95,Bica96,Parker96,Heydari00,Caulet08,Carlson12,Ochsendorf17,Bestenlehner20}. For the rest of the study, we used the model with an age of 10$^{6.6}$ ($\sim$4 Myrs), noting that it is not possible with our tracers to constrain ages younger than this. We also point out that the final results of our study are not affected if younger ages are chosen. 

Based on Figure \ref{fig-OIIIov24_age}, we can also begin to investigate the likelihood of some values of the density and ionization parameter. The models with $\log(n_0)=3$ seem less consistent with the bulk of the data, and only those with a lower \cpdr\ agree with the data. On the other hand, models with low density ($\log(n_0)=1$) and low \cpdr\ ($\leq$50\%) are also not very consistent with most of the data. A deeper investigation is done in the following sections, but the global trend from this figure shows that an intermediate density ($\log(n_0) \sim 2$) is more consistent for the bulk of the data. From Figure \ref{fig-OIIIov24_age}, it is not possible to conclude on the ionization parameter or the \cpdr\ value.

\subsection{Determining the ISM conditions}\label{sect-ISM_conditions}

We then considered the model parameters, \n\ and U. We inspected the ratio \OIIItfratio\, as a function of \Rpdrhii. A high value of \Rpdrhii\ indicates that the emission originating from PDRs is important. As the MIPS 70\mic\, band emission generally traces cooler dust than that of the 24\mic\, band, and is emitted mainly by PDRs in our models, we also compared the ratio \OIIIsvratio\, to \Rpdrhii. 

\subsubsection{Global results for the sample}

\begin{figure}[h!]
\includegraphics[scale=0.4]{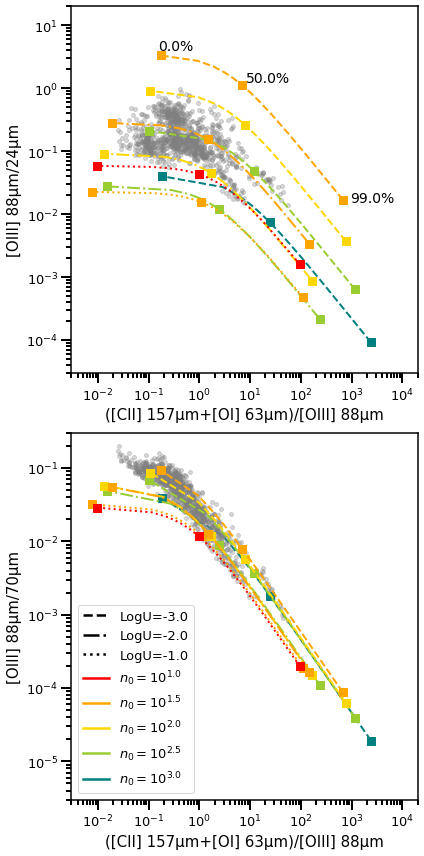}
\caption{Representation of the data of the full sample and models with constant pressure for \specie{O}{iii} 88\mic/MIPS 24\mic\, as a function of \PDRHIIratio\, (top panel), and \specie{O}{iii} 88\mic/MIPS 70\mic\, as a function of \PDRHIIratio\, (lower panel). Each model is represented with a color for the initial density, \n, and a line style for U, and the lines follow the \cpdr, with the markers representing the proportions of 0\%, 50 \% and 99\%. The models displayed are calculated for an age of star-forming burst of 4 Myrs. The gray points are the observed values for each pixel of the regions.}
\label{fig-OIIIovMIPS_CIIovOIII_Pcst_age4Myrs}
\end{figure}

The observed \OIIItfratio\ and \OIIIsvratio\ ratios are represented as functions of \Rpdrhii\ in Figure \ref{fig-OIIIovMIPS_CIIovOIII_Pcst_age4Myrs} with constant pressure models overlaid, for an age of 4 Myrs. We see that the modeled ratios with the PDR proportions agree with the data, with a relatively flatter trend at low PDR proportions (50\% or less), and a decreasing trend for higher proportions (over 50\%). The use of \OIIItfratio\ and \OIIIsvratio\ together also allows us to better constrain the models, and find a value of U and \n\ agreeing with all of the observations. The models that best agree with the entire data set have log(\n)=2.5, log(U)=-3 and various values for \cpdr. Notice that the \cpdr\ fitting the data is below 50\% for the most part. This means that the emission can be modeled with a single connected \HII\ region and PDR model and a covering factor of the PDR between 0 and 1\footnote{A PDR proportion lower than 50\% corresponds to a covering factor between 0 and 1, whereas a PDR proportion higher than 50\% corresponds to a covering factor greater than 1.}. 

We see that the values of the physical parameters seems quite homogeneous for our sample. However, there is a non negligible dispersion of the data, thus we investigated the physical conditions for each region individually.

\subsubsection{Individual regions}\label{sect-individual_regions}

The different regions have been investigated individually (see Figures in Appendix \ref{appdx-regions_plots}). We see for each region that the models that best fit the data are close to the best model found for the whole sample, but differences can be seen from one region to another. We summarized the best fitting models for the total sample and the individual regions in Table \ref{tab-bestmodels_param}.

In addition to this best model, there are other models in our grid that agree with parts of some regions. Those models are indicated in the second column of Table \ref{tab-bestmodels_param}. The possibility for multiple models to agree with part of the data can be understood as our regions are not homogeneous: there are clumps and structures that are spatially resolved in our maps, thus having different physical conditions, that correspond to different models. The best model for each region is then the most representative model of the global conditions in the region, despite possible local variations.

We see that when multiple models agree, an increase of U corresponds to a decrease of \n\ (Table \ref{tab-bestmodels_param}). It is probable that the density we report here corresponds to a rather diffuse ionized gas, that is ubiquitous in our maps. Although there are dense clumps observed in almost all of the regions in our sample, the subregions with the highest densities are hidden by two effects: the dilution of dense clumps by the surrounding lower density medium, due to the resolution and sampling of our data, and the fact that we report the initial densities at the illuminated edge of the cloud, where the actual densities in our models can rise to a few 10$^5$-10$^6$\cmthree\ at high \Av, corresponding to the high-density clumps observed.

\begin{table*}[h!]
\caption{Solutions for constant pressure models that reproduce the full global data set of all regions together as well as the individual regions.}
\begin{center}
\begin{threeparttable}
\begin{tabular}{ccc}
\hline
 & Best model & Other agreeing models\\
 & $\log(U)$, log(\n) & $\log(U)$, log(\n)\\
\hline
Global sample & $-3, 2.5$ & $-2$, [$1.5,2$]\\
 & & $-3$, [$2, 2.5$]\\
30 Doradus & $-3, 2.5$ & $-3$, [$1.5,2$]\\
 & & $-2$, [$1.5,2$]\\
N11B & $-3$, [$2,2.5$] & -  \\
N11C & $-3, 2$ & $-3, 2.5$ \\
N11I & $-3, 2$ & $-3, 2.5$ \\
N44 & $-2, 1.5$ & $-2, 2$ \\
 & & $-3$, [$2.5,3$] \\
N158 & $-3, 2.5$  & $-2, 1.5$\\
N159 & $-3, 2.5$ & $-2$, [$1.5,2$]\\
N160 & $-3, 2.5$ & $-2$, [$1.5,2$]\\
N180 & $-3$, [$2,2.5$] & $-2, 1.5$\\
N66 & $-3, 2$ & - \\
\hline
\end{tabular}
\label{tab-bestmodels_param}
\begin{tablenotes}
\item As the observations trace a wide range in parameter space, other models of our grid can be consistent with each region (although less than our best model), or with a subpart of the region. Those models are indicated in the third column of the Table.
\end{tablenotes}
\end{threeparttable}
\end{center}
\end{table*}

\begin{figure}[h!]
\includegraphics[scale=0.4]{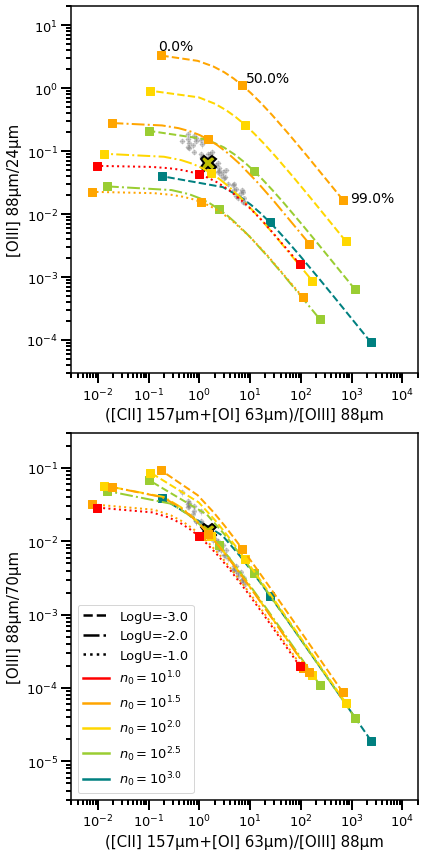}
\caption{Similar to Figure \ref{fig-OIIIovMIPS_CIIovOIII_Pcst_age4Myrs}, but for the region N44.}
\label{fig-OIIIovMIPS_CIIovOIII_Pcst_N44}
\end{figure}

N44 (Figure \ref{fig-OIIIovMIPS_CIIovOIII_Pcst_N44}) is the region that deviates most from the global distribution of the parameters, being the only region with the highest ionization parameter, $\log(U)=-2$, while all of the other regions and the global value of N44 indicate $\log(U)=-3$. N44 also has a lower \n\ value: $\log($\n$) = 1.5$ while the other sources have $\log($\n$)$ values between $2$ and $2.5$. \cite{McLeod19} found that a large bubble (45.7 pc radius) was carved out in the N44 region by the central star cluster, leading to a large zone with very low density ($n \leq$ 100 \cmthree) with multiple regions of star formation triggered around this main bubble, each of them carving out the ISM resulting in different measured densities. Our PACS region is similar to the subregion of triggered star formation N44C for both the position and the size of the map, although our maps are slightly shifted toward the main bubble. Since the very low density medium inside the bubbles cover almost half of our PACS map, this morphological observation may explain the global low density we found, which leads to a higher ionization parameter for a given radius. It is a condition unique to N44, as the other regions in our sample do not show such a large excavation of the ISM. However, we also see from Table \ref{tab-bestmodels_param} and Figure \ref{fig-OIIIovMIPS_CIIovOIII_Pcst_N44}, that other models with with lower U ($\log(U)=-3$) and higher densities ($\log($\n$) = 2.5 to 3$) can be consistent with a subregion of N44, which is in agreement with the density found by \cite{McLeod19} for the N44C region. Here they measure an electron density of 152 $\pm$ 42 cm$^{-3}$ with a much smaller aperture ($\sim$ 8\arcsec) than that of our PACS observations. Thus, best model values are consistent with globally very low density, where the other models with $\log(U)=-3$ and $\log($\n$) = 2.5$ to $3$ agree well with the higher density structures of swept up material in the region.
 
Our best model is selected by picking the model that agrees best with the trend delineated by the observations. However, in some regions, the observations are consistent with different models in our grid. In Figure \ref{fig-OIIIovMIPS_CIIovOIII_Pcst_N44}, the model grids flatten at low values of \Rpdrhii\, with an almost constant \PDRHIIratio, inconsistent with the observations which show continually rising values of \Rpdrhii. In the case of N44, we see on Figure \ref{fig-OIIIovMIPS_CIIovOIII_Pcst_N44}, that models with a lower U and a higher \n\ are more consistent with the observations at low values of \Rpdrhii. In this region, it can be interpreted as the edge of the bubble described in \cite{McLeod19}.

While the PDR tracers, \oi\ and \cii\ are prominently emitted in all of the star-forming regions studied, most regions have quite low proportions of PDR components along the lines of sight throughout their regions (see Figure \ref{fig-PDRproportion_sketch} for visual representation of different PDR proportions). All but 5 regions have \cpdr\ $<$ 12\% throughout their maps. The N11C region shows pixels with \cpdr\ at least 30\% along their lines of sight and only three regions exhibit \cpdr\ $>$ 50\% for some of the pixels of the regions: N44, N11I and N159. Both N159 and N44 show complex morphologies, with arcs, bubbles, and clumps. They also show ongoing star formation, with associated molecular clumps. On the other hand, N11I harbors a more quiescent and extended molecular clump. The higher \cpdr\ values exhibited toward parts of the maps could be linked to the recent stellar formation activity and the birth cloud that may not be totally dispersed. For N11I, it could also be linked to ongoing star formation inside the quiescent molecular cloud. As most of the regions have low \cpdr, this implies that most of the pixels are largely dominated by \HII\, regions, and there is little neutral gas that can be observed along the lines of sight compared to the more prominent ionized gas. On the other hand, pixels where the PDR proportion is higher than 50\% correspond to an excess of PDRs considering a simple model with a connected \HII\, region plus PDR. This may mean that we have multiple clouds along the line of sight, or that we are looking through the external part of a cloud that is spatially resolved. In all of the cases, we see that the PDR proportion varies significantly within an individual region. In the following we thus investigated the spatially resolved distribution of \cpdr, to have a better understanding of how it can be linked to physical interpretation.

\subsection{Insight on the distribution of ionized and neutral gas}\label{sect-PDR_prop}

\begin{figure}[h!]
\includegraphics[scale=0.4]{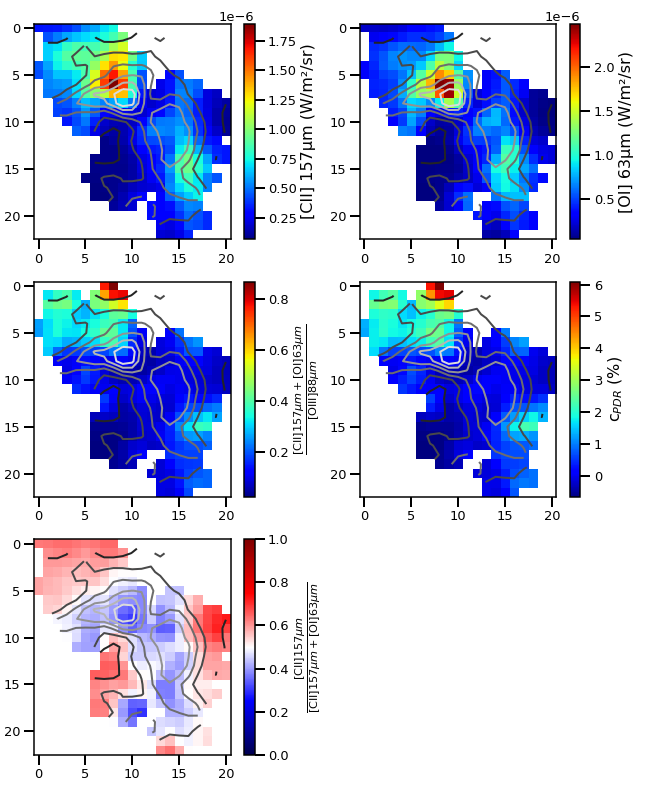}
\caption{The distribution of \cii\ and \oi\ emission (upper left and right panels respectively) in 30 Doradus; the ratio of (\cii+\oi)/\oiii\ (middle left panel), the distribution of \cpdr\ (middle right panel) with the colorbar indicating the percentage of PDR along the line of sight and the ratio \cii/(\cii+\oi) (lower panel). The contours of \oiii\ are overlaid on all the panels for comparison, with the color intensity ranging from  dark gray for lower emission to light gray for higher \oiii\ emission.}
\label{fig-PDRprop_30Dor}
\end{figure}

\begin{figure}[h!]
\includegraphics[scale=0.4]{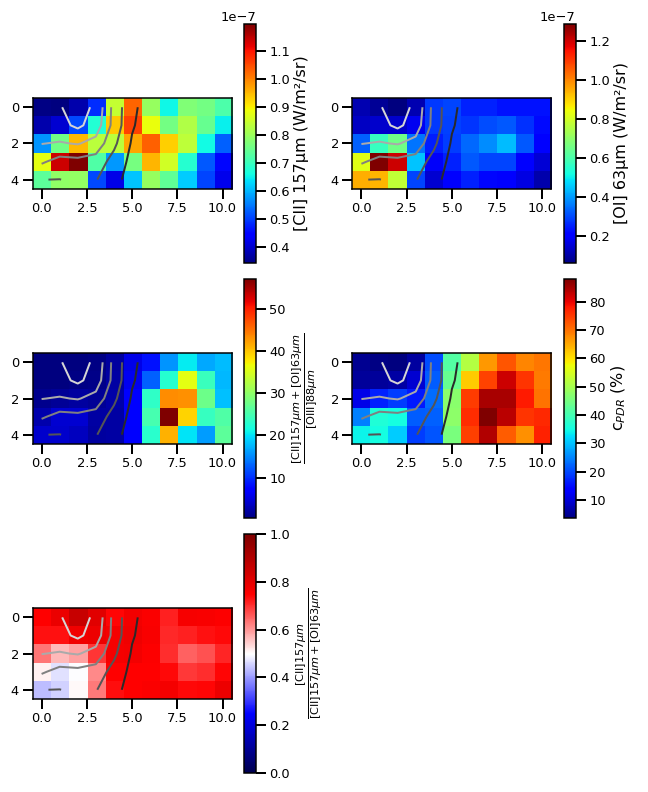}
\caption{Similar to Figure \ref{fig-PDRprop_30Dor}, but for the region N11I.}
\label{fig-PDRprop_N11I}
\end{figure}

For each region, assuming the best fitting model determined in Section \ref{sect-ISM_conditions}, we calculated \cpdr\ needed to match the observed \PDRHIIratio\, ratio (Equation \ref{equ-PDRprop}). Those maps are displayed for 30 Doradus (Figure \ref{fig-PDRprop_30Dor}) and N11I (Figure \ref{fig-PDRprop_N11I}), along with other emission maps and ratio maps for comparison. More region maps are displayed in Appendix \ref{appdx-regions_plots}.

From the comparison of the \cii, \oi\, and \oiii\ emission (Figure \ref{fig-PDRprop_30Dor}; Figure \ref{fig-PDRprop_N11I}), we see that the distribution of the \cpdr\ is, perhaps surprisingly, not closely correlated spatially with that of the PDR tracers, \cii\ and \oi. The spatial distribution of the two PDR tracers is quite similar while they peak somewhat offset from the \oiii\ emission. The lines of sight for the lowest \cpdr\ values are often found in close proximity to the peaks of the \HII\, region tracers. On the other hand, significant \oiii\ emission can be found toward the \cii\ and \oi\ peaks, thus a substantial amount of ionized gas is also present in those pixels and mixed with the PDR tracers. We concluded that strong emission of PDR tracers in one line of sight does not mean that PDRs are dominant along the line of sight. 

We investigated how the physical conditions in the PDRs, in terms of the characteristic properties of the gas traced by \cii\ and \oi\ influence the \cpdr. The cooling of the gas associated with dense and strongly-illumniated PDRs is normally dominated by the \oi\ emission line (e.g., \cite{Tielens85}). On the other hand, \cii\ is the dominant coolant in PDRs that may be less dense, or subjected to lower irradiation. Projection effects could complicate this simple view; it is difficult to access the 3D distribution of matter. This problem arises in particular when trying to determine the density which is a key parameter for a study using \oi. In that vein, \cite{Chevance16} applied PDR models to FIR lines in 30 Doradus, including \cii\ and \oi\ and continuum emission, to constrain the density of the different clouds and investigate the 3D matter distribution. The addition of \oi\ leads to an unexpected distribution of matter around the R136 cluster at the center of the region, showing the strong benefits of including that line. Velocity-resolved observations of the lines can also bring valuable information on the ISM structure \citep[e.g., ][]{Okada15, Requena16, Fahrion17, Okada19, Tarantino21}, helping to disentangle the presence of multiple clumps inside the ISM, to compare the different phases, and to investigate the possibility of self-absorption.

\cite{Okada19} found that \oi\ 63\mic\ shows probable self-absorption features in 30 Doradus and N159 in the LMC, perhaps accounting for $\sim10 - 20\%$ of the total \oi\ emission toward these regions, in some cases, higher than the PACS calibration uncertainties \citep[which is around 10\%, see ][]{Cormier15}. If the \oi\ self-absorption is as high as 20\% in our maps, it  would slightly increase the \PDRHIIratio\ ratio. In our analysis, it may increase the derived \cpdr\ value, but this effect would be small or insignificant and the global results of our study would not be strongly changed.

\begin{figure}[h!]
\centering
\includegraphics[scale=0.5]{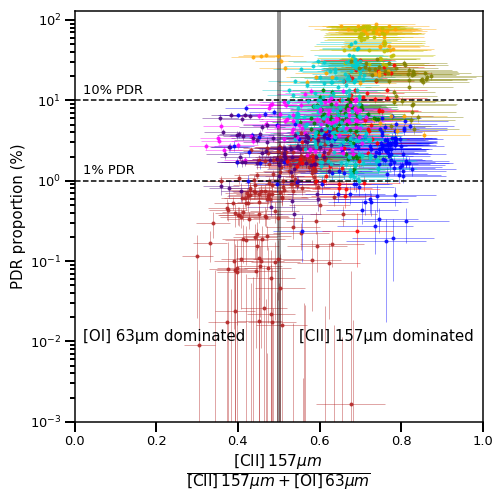}
\caption{The relation between the PDR proportion and the ratio \specie{C}{ii}/(\specie{C}{ii}+\specie{O}{i}), for all of the pixels in all of the regions studied here, with the same region color code as in Figure \ref{fig-OIIIbasic_correlation}. 
The 1\% and 10\% \cpdr\ values are indicated by the two dashed horizontal lines. The vertical gray line separates regions where cooling is dominated by \cii\ (right) and regions where the cooling is dominated by \oi, (left).}
\label{fig-PDRprop_vsdominantline}
\centering
\end{figure}

We compared \cpdr\ maps to the ratio \CIIPDRratio, which we call, \Rciipdr\ (Figures \ref{fig-PDRprop_30Dor}, \ref{fig-PDRprop_N11I} and \ref{fig-app_30Dor} to \ref{fig-app_N66}). We see (Fig. \ref{fig-PDRprop_vsdominantline}) that \Rciipdr\ is correlated with \cpdr. When \Rciipdr\ is a low value, \cpdr\ is also low, and \HII\, regions dominate (see e.g., Fig. \ref{fig-PDRprop_30Dor}). When the ratio is high, \cpdr\ is higher and tends toward PDRs dominating (see e.g., Fig. \ref{fig-PDRprop_N11I} or \ref{fig-app_N44}). The \cii\ line is the major tracer in PDRs that are less dense or less irradiated than those in which the \oi\ line is the dominant tracer, thus it would be tempting to interpret \oi-dominated PDRs as regions that are closer to the illumination source. This interpretation of \Rciipdr\ would be consistent with the \cpdr\ value we derive: when \Rciipdr\ is low, it would indicate that the PDR is close to the ionizing source, which is consistent with the low \cpdr\ value and that the \HII\ region emission dominates along the line of sight. Further from the sources, we see less ionized gas along the line of sight. \cpdr\ increases, and the cooling of PDR gas becomes dominated by \cii\ emission, leading to an increase of \Rciipdr. 

These results show that our approach is useful in investigating the spatial distribution of gas in a 3D sense. It may provide a rough idea of the proximity of the cloud to the ionizing sources, by combining the \oiii, \cii\ and \oi\ lines with the MIPS bands at 24\mic\, and 70\mic.

\subsection{Physical parameters driving the dispersion of the relation between \oiii\ and 24\mic}

We now wish to know how the physical conditions we are studying affect the relation between \oiii\ and 24\mic, and especially their predictive power. As we saw, the solutions for ionization parameters and densities are similar for all of the regions, except for N44. Thus, these parameters do not seem to have a significant effect on the relation and its predictive power. 

\begin{figure}[h!]
\centering
\includegraphics[scale=0.5]{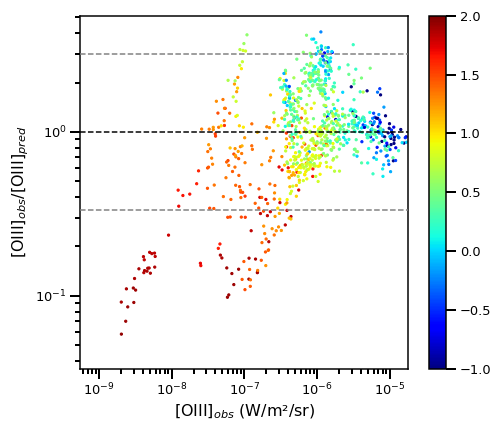}
\caption{The ratio of \specie{O}{iii}$_{obs}$/\specie{O}{iii}$_{pred}$ represented as a function of \specie{O}{iii}$_{obs}$, similar to Figure \ref{fig-OIII_predovobs}, except that the color code corresponds to the log(\cpdr).}
\label{fig-OIII_predobs_PDR}
\centering
\end{figure}

Figure \ref{fig-OIII_predobs_PDR} shows that pixels with \cpdr\ of 50\% or higher have over-predicted values of \oiii, often higher than the observation by a factor of 3 or more. On the other hand, the pixels that show under-predicted \oiii\ values also exhibit some of the lowest PDR proportions found (a few percent or less). Both effects could be explained by the fact that the relation was fitted on the whole sample, including regions where the PDR emission dominates the line of sight. This behavior is not surprising. Indeed, pixels with a higher \cpdr\ are those for which \oiii\ and 24\mic\ emission are less tightly associated, as there is a larger part of 24\mic\ arising from PDRs. This leads to enhanced values of 24\mic\ compared to the \oiii\ emission, and thus to over-prediction of the \oiii\ emission when using the calibration presented here. For pixels with low \cpdr\ values that show under-predicted \oiii, this can be partly linked to the use of data with high \cpdr\ in the fit. The pixels with high \cpdr\ values can influence the global fit, which is thus less accurate for the low \cpdr\ pixels. However, not all of the pixels with low \cpdr\ value show an underestimation of \oiii. Comparing Figure \ref{fig-OIII_predovobs} and Figure \ref{fig-OIII_predobs_PDR}, the pixels with low \cpdr\ values and under-predicted \oiii\ emission mostly come from the regions N11B and N11C. From Table \ref{tab-bestmodels_param}, we see that only models with intermediate density ($\log(n_0) = 2 - 2.5$) are relevant for those two regions, where models with lower density ($\log(n_0) \leq 1.5$) are appropriate for small parts in the other regions. Thus, differences between the physical conditions of the regions can have a significant effect on the prediction. All of those effects make the predictions less accurate for pixels with extreme PDR proportions. 

\begin{figure}[h!]
\centering
\includegraphics[scale=0.5]{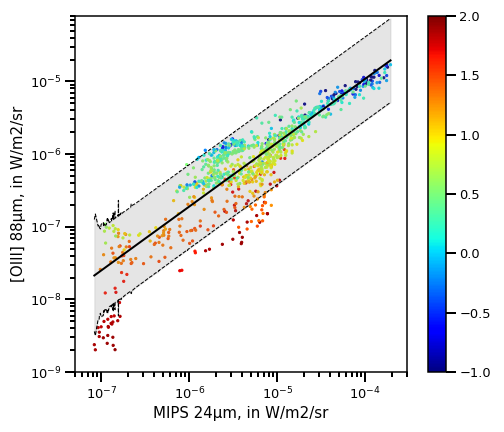}
\caption{Relation between \oiii\ and MIPS 24\mic\, in all of the pixels in all of our star-forming regions for those pixels above 5$\sigma$ in the \oiii\ line. This figure is similar to Figure \ref{fig-OIIIbasic_correlation}, but the points are color-coded with the value of log(\cpdr) here.}
\label{fig-OIIIbasic_correlation_PDR}
\centering
\end{figure}

\begin{figure*}[h!]
\centering
\includegraphics[scale=0.5]{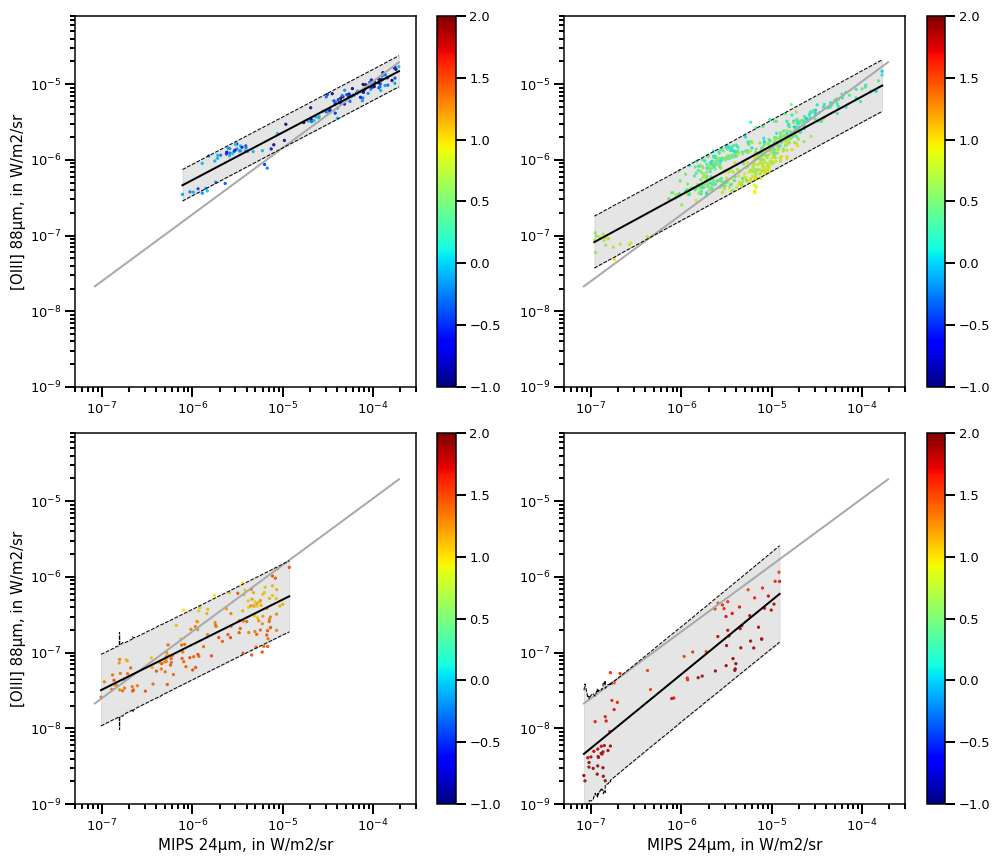}
\caption{Relation between \oiii\ and MIPS 24\mic. The points are color-coded with the value of log(\cpdr). The different panels correspond to the four bins applied on log(\cpdr), namely $[$-1,0$]$ (upper left panel), $[$0,1$]$ (upper right panel), $[$1,1.5$]$ (lower left panel), $[$1.5,2$]$ (lower right panel). The black solid line for each panel correspond to the fitted relation on the bin, the black dashed lines represent the 95\% confidence interval, and the gray solid line is the relation fitted in Figure \ref{fig-OIIIbasic_correlation} (Section \ref{sect-direct_relation_resolved})}
\label{fig-OIIIbasic_correlation_PDR_bins}
\centering
\end{figure*}

We also investigated the effect of PDR proportion on the direct relation between \oiii\ and 24\mic, shown in Figure \ref{fig-OIIIbasic_correlation_PDR}, where we notice some spread of the data relating to the PDR proportion. The pixels with similar \cpdr\ values seem to follow a shallower trend than the global relation, with an offset depending on the value of \cpdr. To test this point, we fit the relation between \oiii\ and 24\mic\, for 4 different bins of PDR proportion with a weighted $\chi^2$. The results from the fitting are presented in Table \ref{tab-fitting_results_PDR}. The data and the relation fitted for the different bins are shown in Figure \ref{fig-OIIIbasic_correlation_PDR_bins}.

\begin{table}[h!]
\caption{Fit parameters of the relation between \oiii\ and 24\mic\, for different bins of log(\cpdr), with the values fit on the full sample noted in the last row, and 1$\sigma$ uncertainty on the slope and intercept. Those fits are represented for each \cpdr\ bin in Figure \ref{fig-OIIIbasic_correlation_PDR_bins}.}
\begin{center}
\begin{tabular}{cccc}
\hline
log(\cpdr) bin & Slope & Intercept & R$^2$\\
\hline
$[$-1,0$]$ & 0.628 $\pm$0.013 & -2.500 $\pm$0.060 & 0.947\\
$[$0,1$]$ & 0.649 $\pm$0.014 & -2.568 $\pm$0.074  & 0.800\\
$[$1,1.5$]$ & 0.595 $\pm$0.033 & -3.324 $\pm$0.194 & 0.687\\
$[$1.5,2$]$ & 0.976 $\pm$0.048 & -1.430 $\pm$0.291 & 0.837\\
all data & 0.879 $\pm$0.012 & -1.446 $\pm$0.065 & 0.824\\
\hline
\end{tabular}
\label{tab-fitting_results_PDR}
\end{center}
\end{table}

Table \ref{tab-fitting_results_PDR} shows that fits to the data of the bins of lower \cpdr\ have a shallower slope than the global fit of Figure \ref{fig-OIIIbasic_correlation} (Section \ref{sect-direct_relation_resolved}). While the slopes for the three lowest bins are consistent with each other within the uncertainties, they are offset from each other, and not consistent with the global relation fitted earlier (see Equation \ref{equ-basic_fit} in Section \ref{sect-direct_relation_resolved}). On the other hand, the relation for the bin of highest \cpdr\ values (log(\cpdr) =1.5 to 2.0) has a slope closer to, but somewhat steeper than that fitted on the full sample. This suggests that the fit on the complete sample is strongly affected by the pixels with high \cpdr, despite their low numbers.

While the calibration is greatly improved by this separation based on the \cpdr, we recommend using the initial calibration carried out on the global sample for general predictive use. Indeed, the calibration for low \cpdr\ may provide a better prediction for the sight lines dominated by the \HII\, region emission, but it would overestimate more often for regions dominated by PDR emission. Thus, using the calibrations based on \cpdr\ bins would require some prior knowledge of the distribution between the phases of the ISM.

\section{Discussion and caveats}\label{sect-discussion}

This new method estimating the distribution of the ISM phases along sight lines in the studied regions, is based on a few tracers only: \oiii\ to trace the ionized gas, \cii\ and \oi\ to trace the PDR gas, and MIPS 24\mic\, and 70\mic\, tracing the warm and cooler dust emission. We first show that 24\mic\ can have a significant contribution from PDR components, not only from the \HII\, regions as is normally assumed in the literature. The possible 24\mic\ emission from PDRs cannot be neglected. Then, we determined some properties of the star-forming regions, such as U and the \n\ at the illuminated edge of the cloud, which is important to constrain the age of the burst of star formation. Our method also gives access to a new parameter, \cpdr, allowing us to investigate the distribution of gas between \HII\, regions and PDRs along the line of sight in a new and original way.

It is always difficult to investigate the distribution of matter, whether it is spatial distribution or contribution from the different phases of the ISM. A number of studies have been carried out during the past decades to try to disentangle the geometry and distribution effects. These include, but are not limited to, multisector models with different densities, ionization parameters or radiation fields \citep{Cormier15}; the combination of radiation-bounded and matter-bounded models \citep{Cormier15,Polles19}; the addition of other sources of excitation beyond optical/UV stellar light, such as X-rays \citep{Lebouteiller17}; as well as the use of a PDR covering factor parameter applied to a standard Cloudy simulation with \HII\, regions and PDRs \citep{Cormier19}. Other works attempted to make 3D simulations of turbulent star-forming clouds with a chemical network and self-consistent physical calculations, sometimes by coupling different codes \citep[see e.g., ][]{Rollig07, Glover11, Glover15, Glover16, Offner14, Accurso17a,Ramos20}, or investigated the geometry and properties of the ISM with velocity-resolved studies \citep[see e.g., ][]{Okada15, Okada19, Requena16, Bisbas19, Fahrion17, Tarantino21}. Despite all of these efforts, there are still degeneracies related to the conditions and geometry of star-forming regions. Detailed observations with increased spatial and spectroscopic resolution indeed reveal very high levels of complexity in galactic ISM regions. It helps to understand the global picture when the different phases are studied as an ensemble. However, the tracers used can raise different issues, like absorption (optical/UV, but also self-absorption in some conditions, such as the \oi\ line discussed above, when viewed through the Milky Way disk, or edge-on galaxies), and extinction effects. Those issues are not significant for our data, but should be investigated before using the methods developed. The possibility that a single tracer can arise from multiple phases, under some galactic environments, such as \cii\ potentially originating in both PDRs and \HII\, regions, also needs to be accounted for.

As it is usually associated with star-forming regions, the 24\mic\, continuum is often used directly as a tracer of star-formation regions and of the SFR (e.g., \citealp{Hao11,Kennicutt12}). However, for studies that require disentangling the different phases of the ISM in star-forming regions, the origin of 24\mic\, emission can also lead to an ambiguous interpretation, as we have seen. It is often implicitly considered to originate from the ionized phase only \citep[e.g., ][]{Malhotra01, Croxall12}, but our study demonstrated that it is not the case, and that the origin of the 24\mic\, emission from PDRs should also be considered for a more accurate interpretation. By comparing the ratios \OIIItfratio\, and \PDRHIIratio, it should be possible to derive a correction factor (which can also depend on other parameters, such as density) to apply to 24\mic\, to separate the relative contribution from \HII\, regions and PDRs.

One of the interests of this work is that our method relies on IR tracers, that are not normally extinguished nor absorbed in most cases. In particular \cii\ and \oiii\ are among the brightest FIR emission lines in galaxies, and are popular tracers being observed in high-redshift galaxies with ALMA. \cii\ observations have been carried out up to a redshift of z$\sim$6 in the large ALPINE sample of \citealp{LeFevre20} and associated studies, \citealp{Carniani20}, or even at z$\sim$7 or higher\citep{Hashimoto20, Bouwens21}. Likewise, the \oiii\ line has also been detected at z$\sim$7 \citep[e.g.,][]{Hashimoto20, Bouwens21}. The rest-frame 24\mic\ band has also been observed for galaxies at different redshifts with \spitz\ and \hers\ telescopes \citep[up to z $\sim 3.5$, see e.g., ][ among others]{Yan07, Oliver12, Liu21, Nagaraj21, Schouws22}, and is accessible with current facilities, such as FIFI-LS \citep{Fischer18}, GREAT \citep{Heyminck12, Risacher16} and HAWC+ \citep{Harper18} on board the Stratospheric Observatory for Infrared Astronomy \citep[SOFIA, ][]{Young12}. In the future, the Fred Young Submillimeter Telescope (FYST) may be able to access 24\mic\ in galaxies near the epoch of reionization.

The origin of the \cii\ line can be uncertain, more so in high metallicity galaxies. It can be emitted by both neutral and ionized gas as it requires 11.3 eV to create \cplus. However, particularly in low metallicity galaxies the contribution of ionized gas to \cii\ is often small (typically less than 30\%, \citealp{Cormier19}). 

Considering the high redshift observations, our new method would thus certainly be convenient to derive a picture of the physical conditions in distant low-metallicity galaxies, where these few tracers already exist. There are, however, additional caveats to keep in mind when using this method to probe PDR and \HII\ region distributions.

First, we assume that in a given pixel the different phases (\HII\ regions, PDRs and neutral gas) have the same physical conditions, in particular the same density at the illuminated edge of the cloud and the same ionization parameter. Considering that the ISM is clumpy, it is not likely that all of the clumps along the line of sight have the same density, or the same illumination conditions. The assumption of constant pressure through all the models can also be questioned. The density profile may be more complex than these simple assumptions, and other effects can disturb the assumed equilibrium (winds, steps in pressure, turbulence, flows, etc.). 

Another caveat is that our models are computed up to \Av\ of 2.5. This can lead to very high densities (up to a few 10$^6$\cmthree) in the PDR, which can induce strong \cii\ and \oi\ emission. It is probable that not all of the PDRs attain those high depths, and an investigation of different final cuts of the model have to be taken into account eventually.

Bringing more diagnostic tracers into the modeling scheme may also help to refine the results and make them more robust. \spitz\ has observed numerous MIR \HII\, region tracers such as \specie{Ne}{ii} and \specie{Ne}{iii}, \specie{Si}{ii}, iron lines and molecular hydrogen lines in some regions, potentially helping to further constrain the conditions in PDRs and \HII\, regions mixed within the telescope beam. {\it JWST} will sure bring us a rich array of these tracers. Additionally, velocity resolved observations to separate different clumps and phases, would also bring further information on the phase distributions. Despite all of these challenges, the method presented here brings a new and simple angle to study and disentangle the emission from the ionized and neutral gas in star-forming regions, with the possibility of application to other unresolved low metallicity galaxies, even at high redshift.

\section{Summary}\label{sect-summary}

We examined the relation between the \hers\ \specie{O}{iii} 88\mic\ and the \spitz\ 24\mic\, continuum band for a sample of star-forming regions that have been mapped in the Magellanic Clouds, using a novel and simple method to spatially disentangle different phases along the line of sight. We also compared our spatially-resolved results with those of the unresolved observations of the \hers\ Dwarf Galaxy Survey. Out findings include the following:
\begin{itemize}

\item The \oiii\ and \spitz\ 24\mic\, continuum show a clear correlation covering three orders of magnitude for 24\mic\, and four orders of magnitude for \oiii. We find, in surface brightness:

\begin{equation}
\log(\Sigma_{\rm [OIII]\,88}) = 0.879 (\pm 0.012) \times \log(\Sigma_{\rm 24}) -1.446 (\pm 0.065),
\end{equation}
\noindent with ($\Sigma_{\rm [OIII]\,88}$ and $\Sigma_{\rm 24}$ in \unitsurf)

\noindent and with \loiii\ and \lcont\ in luminosity (\lsol): 
\begin{equation}
\log(L_{\rm [OIII]\,88}) =  0.948 (\pm 0.086) \times \log(L_{\rm 24}) -0.622 (\pm 0.360).
\end{equation}

\item To investigate the local conditions driving this relation, we compared our observations with a grid of 1D Cloudy models containing an \HII\ region component and a PDR component, and vary the stellar age, the initial density (at the illuminated edge of the cloud), \n\ and the ionization parameter, U. We also created a new parameter, the proportion of PDR, \cpdr, to determine the proportion of emission arising from PDRs and that with an origin in \HII\ regions along each line of sight.\\

\item We used the ratio \PDRHIIratio\, as a proxy for the ratio of PDR vs. \HII\, region emission, and compared it to the \OIIItfratio. The use of \OIIItfratio\ and \oiii/70\mic\ together allows us to constrain the models most efficiently.  \\

\item While the 24\mic\, continuum is often considered to trace dust specifically in \HII\, regions, we showed that it can also arise in PDRs, at least in low metallicity environments investigated here. Thus its use as a probe of \HII\ regions only should be treated with caution.\\

\item The PDR proportion, \cpdr,  is spatially anticorrelated with the \oiii\ line emission, emitted only by \HII\, regions, and not necessarily correlated with the two PDR tracers, \cii\ and \oi. This suggests that \cpdr\ is not tracing the simple presence of PDRs. Instead, the \cpdr\ parameter indicates whether PDRs or \HII\ regions are the dominant phase along the line of sight. \\

\item We found that \cpdr\ correlates with the ratio \specie{C}{ii}/(\specie{C}{ii}+\specie{O}{i}). This suggests that regions of high \cpdr\ values, where PDRs dominate the line of sight emission, originate in environments of relatively lower illumination, thus further away from the young stars that ionize their surroundings.\\
\end{itemize}
We demonstrated the predictive power of \oiii\ and 24\mic\ for both integrated and spatially resolved Magellanic Cloud regions and for the unresolved DGS data.  Areas where the predicted \oiii\  is significantly higher than the actual observations are dominated by PDRs, with higher \cpdr\ values. On the other hand, areas with the lower \cpdr\ values give predictions closely resembling the observations (observed and predicted values agree within a factor 3 for 90\% of the data), although the predicted values have to be taken with caution for very low \oiii\ fluxes ($\leq$ 3 $\times$ 10$^{-8}$ \unitsurf), and for places where the ionized gas may not be dominant.

\section*{Acknowledgements}
We would like to thank the anonymous referee for useful comments that greatly helped to improve the quality of the paper.

This work is based in part on Herschel, which is an ESA space observatory with science instruments provided by European-led Principal Investigator consortia and with important participation from NASA. This work is based in part on observations made with the Spitzer Space Telescope, which was operated by the Jet Propulsion Laboratory, California Institute of Technology under a contract with NASA.


\bibliographystyle{aa} 
\bibliography{allbiblio}

\begin{appendix}
\section{Analytic examples of the modeling strategy}\label{appdx-model_strat}

Following on the description of the modeling strategy described in Section \ref{sect-models}, we elaborate further on other examples of mixing different proportions of \HII\, regions plus PDR Cloudy models, related to the illustrations of Figure \ref{fig-PDRproportion_sketch} which correspond to other physical configurations along the line of sight.

We rewrite Equation \ref{equ-general_intensity} in terms of I$_{\rm single}$, which denotes a single, full Cloudy model. We then scale the full single Cloudy model along with the fraction of PDR intensity (I$_{\rm PDR}$) necessary to match the balance of \HII\, region plus PDR.

\subsubsection{Example: 90\% \HII\, region and 10\% PDR}
 
For a mixed line of sight dominated by an \HII\, region component, Equation \ref{equ-general_intensity} gives:

\begin{equation}
\left\{
\begin{array}{ccl}
I_{\rm single} & = & I_{\rm HII+PDR}\\
0.9 \, I_{\rm H\,II} + 0.1 \, I_{\rm PDR} & = & 0.9 \, I_{\rm H\,II} + 0.9 \, I_{\rm PDR} - 0.8 \, I_{\rm PDR}\\
 & = & 0.9 \, I_{\rm single} -  0.8 \, I_{\rm PDR}\\
 & = & 0.9 \left( I_{\rm single} -  \frac{0.8}{0.9} \times I_{\rm PDR}\right),
\end{array}
\right. ,
\label{equ-PDR10}
\end{equation}

In Equation \ref{equ-PDR10}, we see that a proportion of 10\% of PDR can be simulated as a full Cloudy model with part of the PDR suppressed, assuming all of the modeled ISM have identical conditions (same ionizing sources, same initial density and same initial ionization parameter). The physical interpretation of a single Cloudy model, with part of the PDR removed, may represent a matter-bounded region, where the PDR is decreased due to lack of gas and dust. 

\subsubsection{Example: 50\% \HII\, region and 50\% PDR}

For an equal mix of \HII\, region and PDR, Equation \ref{equ-general_intensity} gives:

\begin{equation}
\begin{array}{ccl}
0.5 \, I_{\rm H\,II} + 0.5 \, I_{\rm PDR} & = & 0.5 \, I_{\rm single}.
\end{array}
\label{equ-PDR50}
\end{equation}

We see in Equation \ref{equ-PDR50} that a mixture with 50\% of PDR (Figure \ref{fig-PDRproportion_sketch}) corresponds to a single Cloudy model, with all of the PDR emission taken into account, still under the assumption that the \HII\, region and the PDR component have the same initial density and ionization parameter. We highlight the point that 50\% of PDR does not correspond to a medium with half of the volume being filled by PDR, but corresponds to a mix that can be modeled by a Cloudy model with a full PDR layer. This configuration can be physically interpreted as lines of sight through the different zones of a star-forming cloud with a young star cluster, and may represent a radiation-bounded  cloud. 

\subsubsection{Example: 10\% \HII\, region and 90\% PDR}

For a mix dominated by PDR region, Equation \ref{equ-general_intensity} gives:

\begin{equation}
\begin{array}{ccl}
0.1 \, I_{\rm H\,II} + 0.9 \, I_{\rm PDR} & = & 0.1 \, I_{\rm H\,II} + 0.1 \, I_{\rm PDR} + 0.8 \, I_{\rm PDR}\\
 & = & 0.1 \, I_{\rm single} +  0.8 \, I_{\rm PDR}\\
 & = & 0.1 \left(I_{\rm single} + 8 \, I_{\rm PDR} \right),
\end{array}
\label{equ-PDR90}
\end{equation}

In Equation \ref{equ-PDR90}, we see that 90\% of PDR in the mixture can be represented by a full Cloudy model with a further addition of eight times more PDR in this case. Those PDRs are connected to \HII\, regions lying outside the line of sight. Such a model also represents a radiation-bounded case along the corresponding line of sight between the stellar source and the cloud. Physically, this configuration can not be modeled by a single Cloudy model, as it would overestimate the emission of tracers emitted by the \HII\, region. 

\section{Tests on the strength of the fit}\label{appdx-test_fit}

We investigate the robustness of the relation between \specie{O}{iii} 88\mic\, and MIPS 24\mic, studied in Section \ref{sect-direct_relation_resolved} when changing various aspects of the analysis: the pixel size, the method used for the fitting, the correction for point sources and the resolution at which we analyze the maps.

\subsection{Test of the fitting method}

Here we compare the effect of the fitting method of the parameters of the linear relation of \oiii\ and 24\mic\ discussed in Section \ref{sect-direct_relation}. A simple, non weighted $\chi^2$, method is made to investigate the uncertainties due to the fitting method. Another method uses a Monte Carlo simulation to calculate the uncertainties, which can then be considered to be directly linked to the observation uncertainties. Finally, the weighted $\chi^2$ used in Section \ref{sect-direct_relation_resolved},  takes into account uncertainties from the data and the method of fit. The fitted slopes and intercepts for each method, and their uncertainties, are shown in Table \ref{tab-method_fit_params}. 

\begin{figure}
\begin{center}
\includegraphics[scale=0.7]{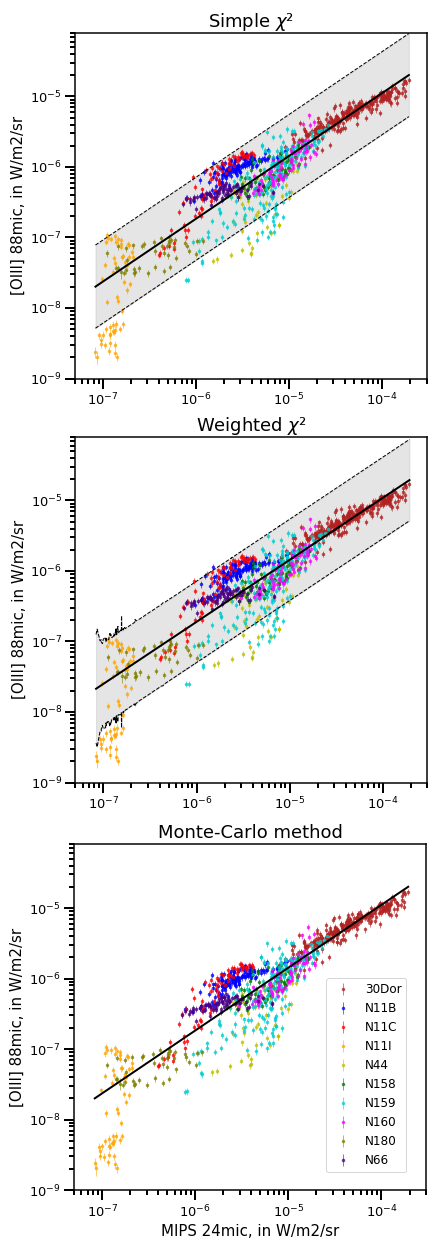}
\caption{Comparison of the linear fits between \oiii\ and 24\mic\ for the three fitting methods. The 95\% confidence intervals are represented with dashed lines and gray shading for the simple and weighted $\chi^2$ methods. The confidence interval is not determined for the Monte Carlo method. The upper right panel, corresponding to the weighted $\chi^2$, is identical to Figure \ref{fig-OIIIbasic_correlation} in Section \ref{sect-direct_relation_resolved}, showing the fit of the correlation.}
\label{fig-OIII_correlation_method}
\end{center}
\end{figure}

The Monte Carlo simulation used the observed sample as a base to run 100 simulated samples, with the same number of pixels, each varying inside the limits of 1$\sigma$, following a normal distribution. The fit is then calculated with a nonweighted $\chi^2$ on each of the simulated samples, and the final value for the parameters are the median from all of the resulting fittings, with the median absolute deviation being the uncertainty. We added the value of the R$^2$ coefficient of the fitted linear regression, to estimate the difference in fitting accuracy. The three fitted relations are illustrated with the data in Figure \ref{fig-OIII_correlation_method}.

\begin{table}[h!]
\caption{Parameters for the different fitting methods of the relationship between \oiii\ and 24\mic\ (linear in log space), discussed in Section \ref{sect-direct_relation}, 1$\sigma$ uncertainties, and R$^2$ parameter as an indication of the goodness-of-fit of the regression model.}
\begin{center}
\begin{tabular}{cccc}
\hline
Method & Slope & Intercept & R$^2$\\
\hline
Weighted $\chi^2$ & 0.879$\pm$0.012 & -1.57$\pm$0.068 & 0.824\\
Simple $\chi^2$ & 0.891$\pm$0.013 & -1.386$\pm$0.066 & 0.826\\
Monte Carlo & 0.892$\pm$0.002 & -1.388$\pm$0.011 & 0.894\\
\hline
\end{tabular}
\end{center}
\label{tab-method_fit_params}
\end{table}

The parameters fitted with different methods are very similar and roughly agreeing within the uncertainties (Table \ref{tab-method_fit_params}). We also see that the R$^2$ values are high for all the methods, between 80 and 90\%, which correspond to a high probabilities of a linear positive correlation. The different methods show very similar relations when compared to the fitted data (Figure \ref{fig-OIII_correlation_method}).
\par The uncertainties derived by the two $\chi^2$ methods are almost identical, and much larger than the uncertainties derived by the Monte Carlo method. It indicates that the uncertainties derived by the fitting procedure and due to the spread of the data are dominant compared to the effect of observation uncertainties. 
\par We can see that the R$^2$ parameters are very similar for the two $\chi^2$ methods, but it is a bit higher for the Monte Carlo simulation, indicating that the Monte-Carlo fit is a bit more representative of the data, but this effect is very small, as the difference between the R$^2$ values is smaller than 0.1.

We can conclude from this test that the different methods give similar results, both on the fitted value and the accuracy of the fit. The comparison of the nonweighted $\chi^2$ and the Monte Carlo method shows that the uncertainties on the fitted parameters driven by the fitting method are much larger than the uncertainties driven by data uncertainties. The weighted $\chi^2$, which is the method finally adopted for our study, has the advantage of taking into account both types of uncertainties, although the uncertainties driven by the fitting method are dominant. 

\subsection{Test on the resolution and sampling of the maps}

\begin{figure}[h!]
\begin{center}
\includegraphics[scale=0.7]{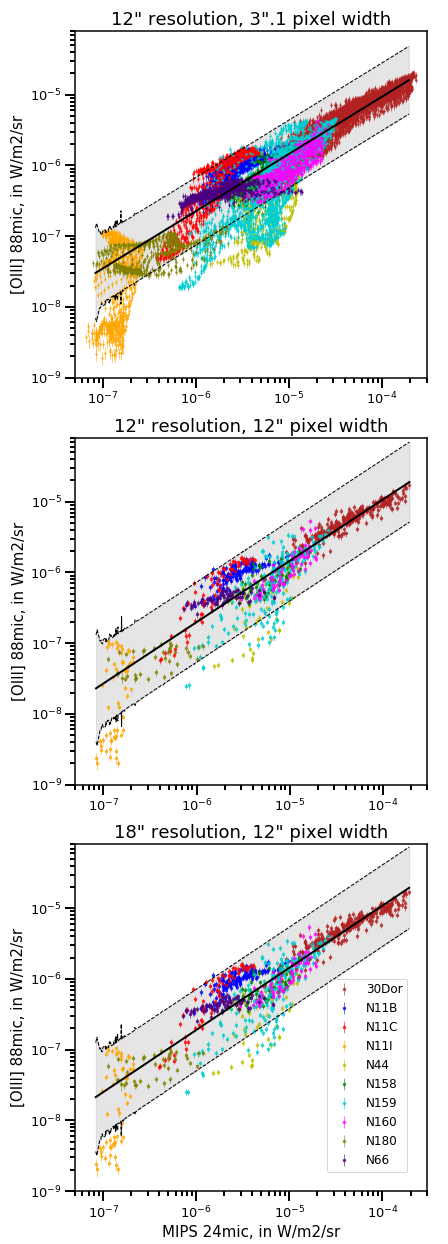}
\caption{The fitted linear relation and the corresponding data for the different resolution and sampling of the maps. The black solid line is the fit, the dashed black lines and gray filling illustrate the 95\% confidence interval. The upper right panel, corresponding to 12" resolution and a pixel size of 12", is identical to the Figure \ref{fig-OIIIbasic_correlation} in Section \ref{sect-direct_relation_resolved}, showing the fit of the correlation.}
\label{fig-OIII_correlation_pxwidth}
\end{center}
\end{figure}

We compare different spatial resolutions and pixel sampling sizes of the maps, to investigate the potential effects it can have on the fit. We use a weighted $\chi^2$ method to be consistent with that used in Section \ref{sect-direct_relation_resolved}, and we compare two different resolutions with the same sampling, and then two different samplings with the same resolution. The resolutions we use are 12", to match the PACS emission lines, and 18" resolution, to match the MIPS 70\mic. For both resolutions, we resample the maps to 12" pixel width, to avoid oversampling of the beam. For the spatial sampling, we compare our final pixel size of 12" to the native sampling of PACS spectroscopy, 3".1. It also allows us to investigate the effect of oversampling the beam, as we use 12" resolution for the tests on sampling size (Figure \ref{fig-OIII_correlation_pxwidth}).

\begin{table}[h!]
\caption{Parameters for the different resolutions and sampling sizes of the maps (Figure \ref{fig-OIII_correlation_pxwidth}), 1$\sigma$ uncertainties, and R$^2$ parameter as an indication of the goodness-of-fit of the regression model.}
\begin{center}
\begin{tabular}{cccc}
\hline
Resolution, pixel size & Slope & Intercept & R$^2$ \\
(arcsec, arcsec) & & & \\
\hline
12, 3.1 & 0.810$\pm$0.009 & -1.788$\pm$0.050 & 0.801\\
12, 12 & 0.867$\pm$0.012 & -1.503$\pm$0.063 & 0.821\\
18, 12 & 0.879$\pm$0.012 & -1.446$\pm$0.065 & 0.824\\
\hline
\end{tabular}
\end{center}
\label{tab-pxwidth_fit_params}
\end{table}

The different resolutions and sampling sizes of the maps return very similar results for the fits of the parameters of the relation (Table \ref{tab-pxwidth_fit_params}), which roughly agree within the uncertainties. The R$^2$ parameters are also very similar for each of the data settings.

The derived uncertainties seem to be affected by the pixel width. For the largest pixel size (12"), the uncertainties on the fitted parameters are 3 times larger than those for the larger pixel width (3".1). Thus, for comparable dispersion, the fit can be more precise where it is carried out on smaller pixels.

While the fits from Figure \ref{fig-OIII_correlation_pxwidth} (Table \ref{tab-pxwidth_fit_params}) do not differ much, the confidence intervals are affected by both the resolution and the pixel size: at the low resolution and low pixel width (12" resolution and 3".1 pixel width), the confidence intervals are highly irregular, especially at the low brightness end of the relation, but they become more regular when increasing the pixel size (12" resolution and 12" pixel width) and when increasing the beam size (18" resolution and 12" pixel width). It corresponds to a smoothing of the data.

We can conclude from this test that the effect of resolution and pixel size is negligible on the fitted parameters of the relation between \specie{O}{iii} 88\mic\, and MIPS 24\mic, even when the beam is oversampled. Although using a larger number of pixels, leads to a somewhat better precision of the fitted values.

\subsection{Test on method for point source correction}

The MIPS 24\mic\, continuum band contains emission from stellar point sources as well as dust emission. 
We investigate the effect of the point source removal, and the effect of the method used, on the fitting of the relation between \oiii\ and MIPS 24\mic.

We developed a dedicated software to extract the emission of the point sources from the maps. The position of the point sources are taken from catalogs. A point source is considered to be a 2D Gaussian. The emission from each point of the map is then decomposed in two components: the background, and the contribution from the point source. The map is divided into tiles and the background is considered to be flat on each tile. The size of the tiles depends on the size of the total map and the spread of the point source emission. The tiles roughly correspond to the 12" pixel except for the 30 Doradus region. The background value is estimated on each tile, in order to fit the 2D Gaussian function for each point source referenced. The software produces three maps as outputs: the tiled background map, a "zoom" background map, which is based on the tiled background and reprocessed to improve the resolution, and a map of the point source emission, at the resolution of the initial map. This point source map is subtracted from the initial map, to retrieve the extended, diffuse emission only. This method comes with some caveats, especially in the case of overestimation of the point source emission. It can lead to some artifacts, or some pixels with negative emission, which are artificially set to zero, but a detailed examination of our sample showed that this type of artifact is not present or is negligible in the studied regions.  

\begin{figure*}
\begin{center}
\includegraphics[scale=0.6]{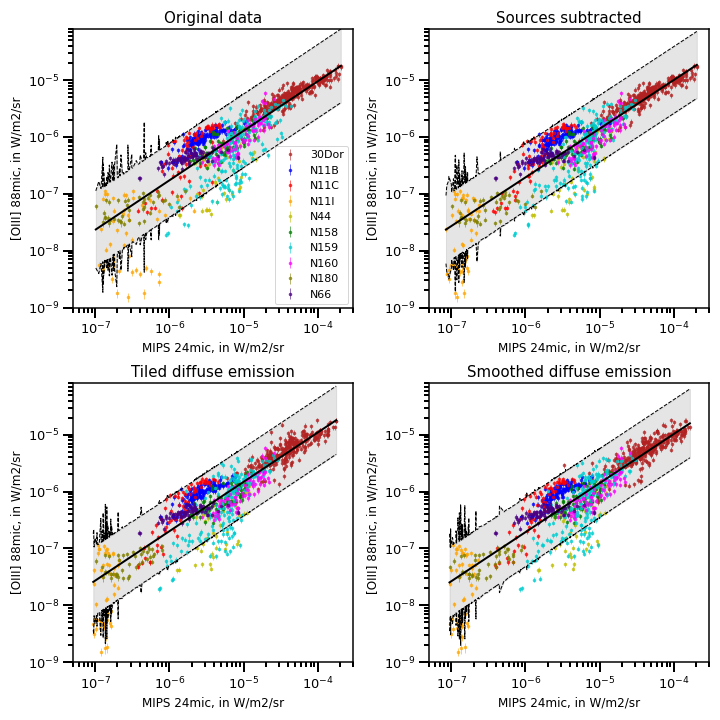}
\caption{Results of the fit with different treatment of point-sources in the 24\mic\ maps. The original MIPS 24\mic\, maps (upper left panel), tiled (lower left panel) and smoothed background (lower right panel) calculated by the software extracting the point sources, and the original maps with emission from the extracted point sources subtracted (upper right panel). The two first maps of the diffuse emission are outputs from the software, where the last one is the software estimation of the point sources emission subtracted from the initial map. The black solid line represents the fitted relation between the \oiii\ emission line and the 24\mic\, band and the black dashed lines and the gray shading indicate the 95\% confidence intervals. Fitted parameters for the different cases are in Table \ref{tab-bkg_fit_params}}
\label{fig-OIII_correlation_ptsrc}
\end{center}
\end{figure*}

In order to study the effect of the method used to correct maps from point source emission, we fitted the relation between \oiii\ and MIPS 24\mic\, by using four 24\mic\, emission maps: the original MIPS 24\mic\, maps, convolved and resampled, without removing the point sources (later called original data), the two types of diffuse emission maps created by the software, namely the tiled background and zoomed background, and the original data with the point source maps subtracted, which is referred to as point source subtracted data or simply subtracted data. In region N160, the 24\mic\, data show a large saturated zone. The saturated pixels were masked, and when trying to apply a point source correction, no point source was found in nonsaturated pixels. Thus the N160 24\mic\, masked map is used in all of the four procedures, without attempts to correct from point source emission. Table \ref{tab-bkg_fit_params} summarizes the parameters fitted for the four 24\mic\, maps, and the results are illustrated in Figure \ref{fig-OIII_correlation_ptsrc}.

\begin{table}
\caption{Parameters for the different treatments of the map, 1$\sigma$ uncertainties, and R$^2$ parameter as an indication of the goodness-of-fit of the regression model.}
\begin{center}
\begin{threeparttable}
\begin{tabular}{cccc}
\hline
MIPS data & Slope & Intercept & R$^2$\\
\hline
Original & 0.872$\pm$0.015 & -1.534$\pm$0.077 & 0.763\\
Tiled diffuse & 0.869$\pm$0.013 & -1.484$\pm$0.071 & 0.797\\
emission & & & \\
Zoomed diffuse & 0.864$\pm$0.014 & -1.532$\pm$0.071 & 0.793\\
emission & & & \\
Point sources & 0.879$\pm$0.012 & -1.446$\pm$0.065 & 0.824\\
subtracted & & & \\
\hline
\end{tabular}
\label{tab-bkg_fit_params}
\begin{tablenotes}
\item "Original" is for noncorrected data, "tiled" and "zoomed diffuse emission" are the two calculation of the diffuse emission from the software used to extract the point sources, and "point source subtracted" is for the original MIPS 24\mic\, maps with point sources subtracted.
\end{tablenotes}
\end{threeparttable}
\end{center}
\end{table}

We notice that the global dispersion is not affected by the method used to correct the 24\mic\,maps from point source emission (Figure \ref{fig-OIII_correlation_ptsrc}), indicating that it is linked to the discrepancies between the different regions, as we have already noticed in Section \ref{sect-direct_relation_resolved}. However, the internal dispersion for each region is affected by the removal of point sources. The two cases of tiled and zoomed background emission reduce the spread for some regions, like N11I or N159, for the MIPS 24\mic\, data (roughly from 1 dex or a bit more to 0.5 dex); it naturally leaves untouched the spread in \oiii\ emission. Some regions, on the contrary, exhibit a spread a bit larger with the tiles and zoomed background than with the original MIPS 24\mic\, data, such as 30 Doradus. This increase of the spread is probably due to the tiling of the diffuse emission used to calculate the point source emission, and thus is identical for the two diffuse emission calculations. The effect is most important for the 30 Doradus region, probably due to the fact that this is the only region for which the tiling is done on a scale much larger than the resampling of the maps at 12" pixel width.

The diffuse emission obtained by subtracting the point source emission from the original maps also does not influence the global spread, but only the spread of each individual region. In this case, we observe the same reduction of the spread for the regions affected by the other point source correction methods, but there is no increase of the spread in the other regions.

Removing the point source emission has little effect on the fitted parameters, independent of the method used (Table \ref{tab-bkg_fit_params}). The slopes and intercepts are very similar, and completely agreeing within the uncertainties. The R$^2$ parameter is somewhat affected: the method of original data, with no point sources subtracted, gives the least accurate fit, the method using two different diffuse emission calculations gives a result a bit more accurate, and the best fitted relation is achieved for the maps with the point source emission subtracted from the original maps. 

From this test, we conclude that taking into account the contamination of the 24\mic\, emission by point source emission is important, especially to investigate the behavior and conditions in the star-forming regions of the individual regions, but it does not influence much the fitting of the parameters of the overall relation between \oiii\ emission and MIPS 24\mic, nor the uncertainties on the fitted parameters. To avoid an increase of the spread in some individual regions, and the loss of spatial resolution due to the tiling used to calculate the point source emission, we use the MIPS 24\mic\, data with point sources subtracted.

\newpage
\section{Maps and specific plots for the individual regions}\label{appdx-regions_plots}

\begin{figure*}[h!]
\begin{center}
\includegraphics[scale=0.3]{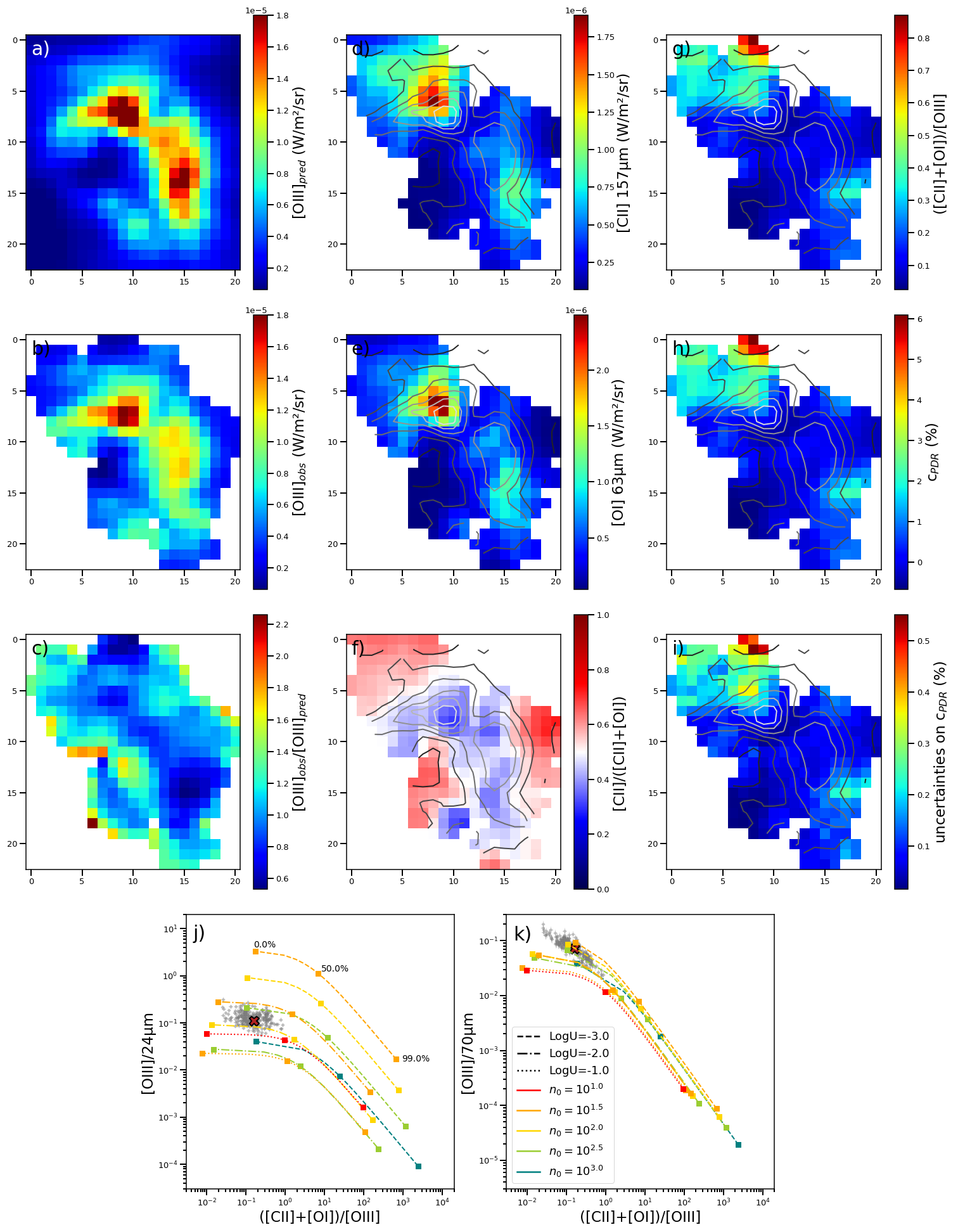}
\caption{For the region 30 Dor, the predicted map of \oiii\ given the observed 24\mic\, (panel a) is compared to the actual map of \oiii\ observed with \hers/PACS (panel b), and the ratio of observed over predicted emission (panel c). The distribution of \cii\ and \oi\ emission (panels d and e respectively) are overlaid with the \oiii\ line contours. The ratio of \Rciipdr\ (panel f), \Rpdrhii\ (panel g) and \cpdr\ with associated uncertainties (panel h and i respectively) are also presented with the \oiii\ line contours. The observed ratios \oiii/24\mic\ and \oiii/70\mic\ presented as a function of \Rpdrhii\ are compared to our grid of models (panelj and k respectively). The large cross is the value integrated on the global region.}
\label{fig-app_30Dor}
\end{center}
\end{figure*}

\newpage

\begin{figure*}[h!]
\begin{center}
\includegraphics[scale=0.3]{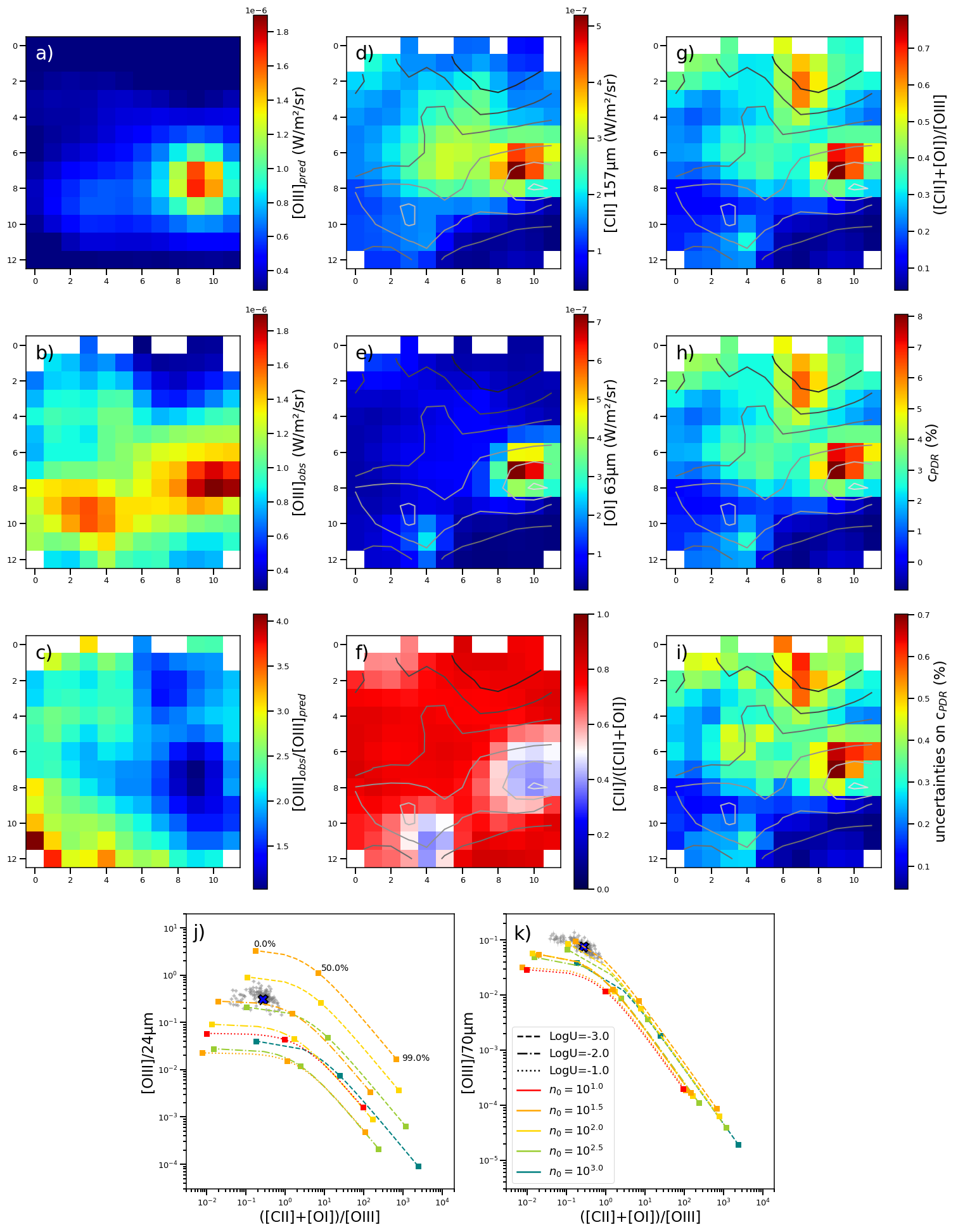}
\caption{Similar to Figure \ref{fig-app_30Dor}, but for the region N11B.}
\label{fig-app_N11B}
\end{center}
\end{figure*}

\newpage

\begin{figure*}[h!]
\begin{center}
\includegraphics[scale=0.3]{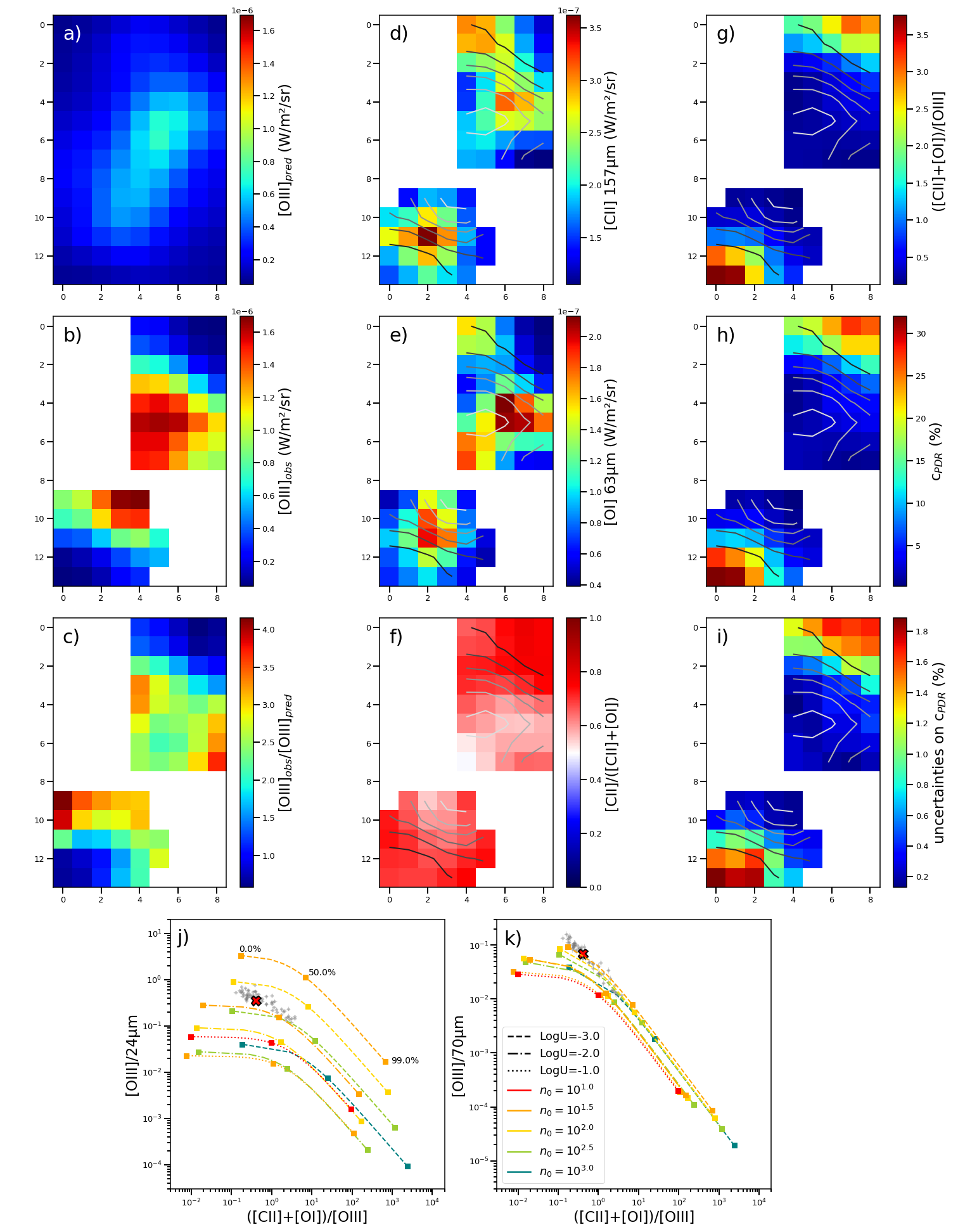}
\caption{Similar to Figure \ref{fig-app_30Dor}, but for the region N11C.}
\label{fig-app_N11C}
\end{center}
\end{figure*}

\newpage

\begin{figure*}[h!]
\begin{center}
\includegraphics[scale=0.3]{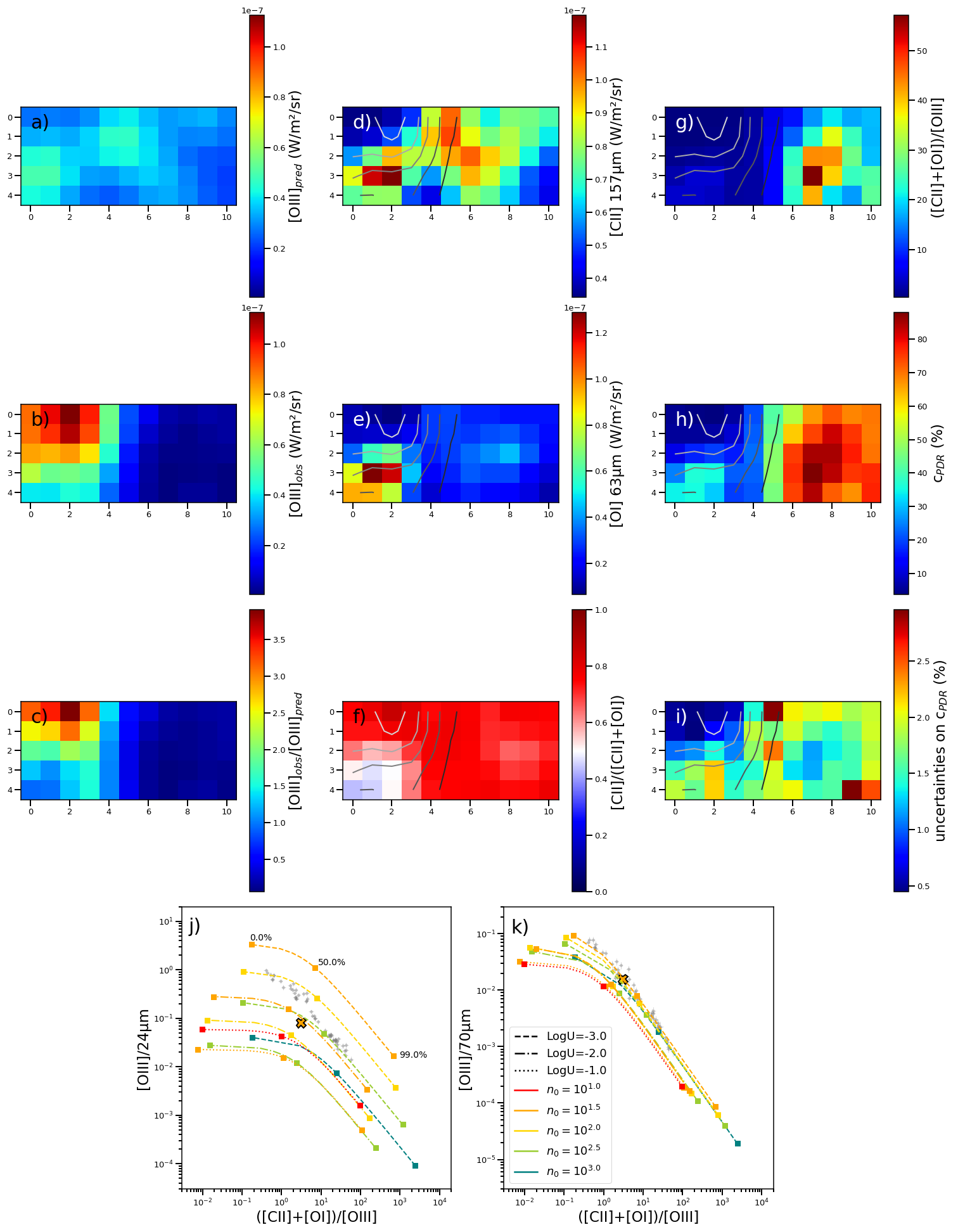}
\caption{Similar to Figure \ref{fig-app_30Dor}, but for the region N11I.}
\label{fig-app_N11I}
\end{center}
\end{figure*}

\newpage

\begin{figure*}[h!]
\begin{center}
\includegraphics[scale=0.3]{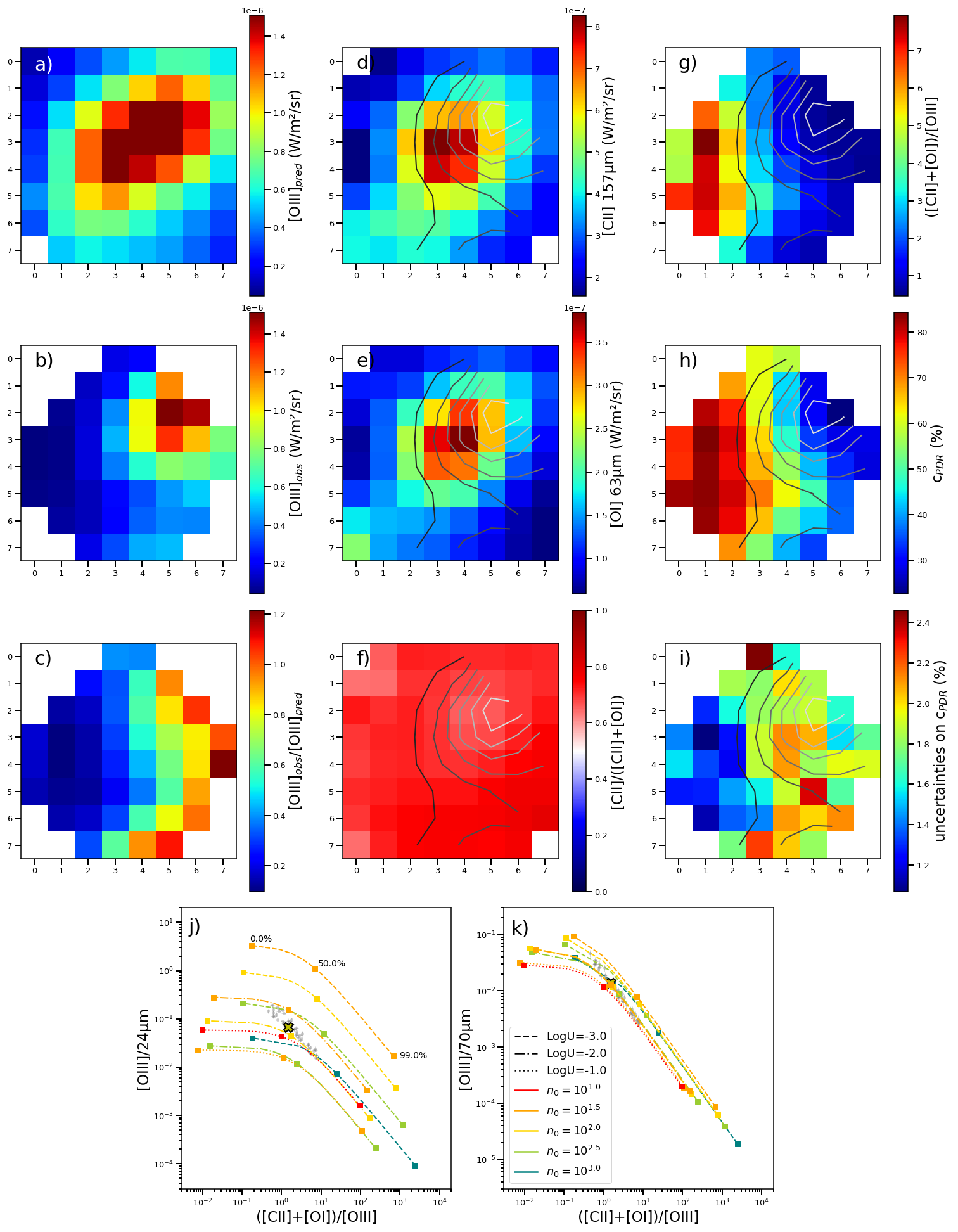}
\caption{Similar to Figure \ref{fig-app_30Dor}, but for the region N44.}
\label{fig-app_N44}
\end{center}
\end{figure*}

\newpage

\begin{figure*}[h!]
\begin{center}
\includegraphics[scale=0.3]{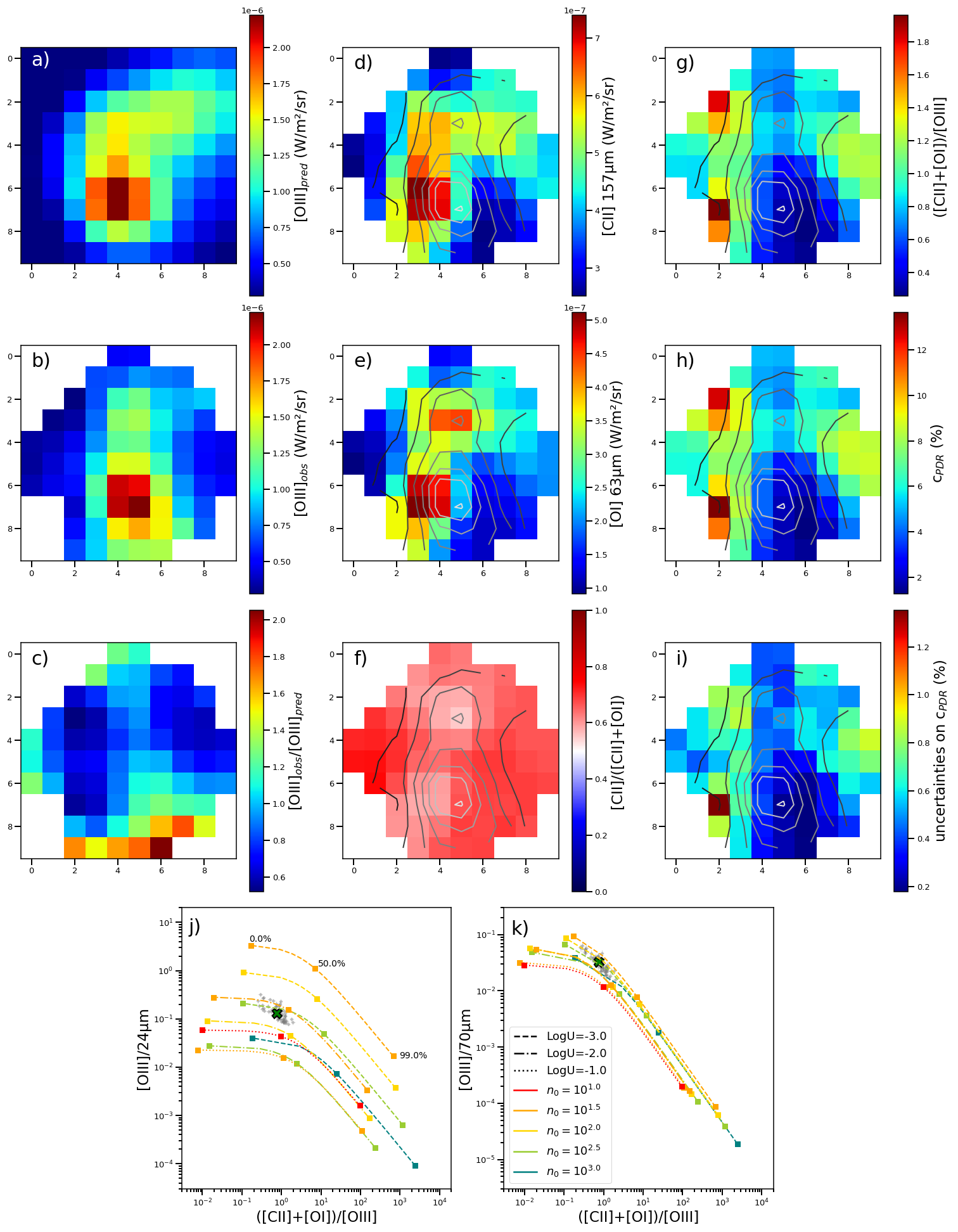}
\caption{Similar to Figure \ref{fig-app_30Dor}, but for the region N158.}
\label{fig-app_N158}
\end{center}
\end{figure*}

\newpage

\begin{figure*}[h!]
\begin{center}
\includegraphics[scale=0.3]{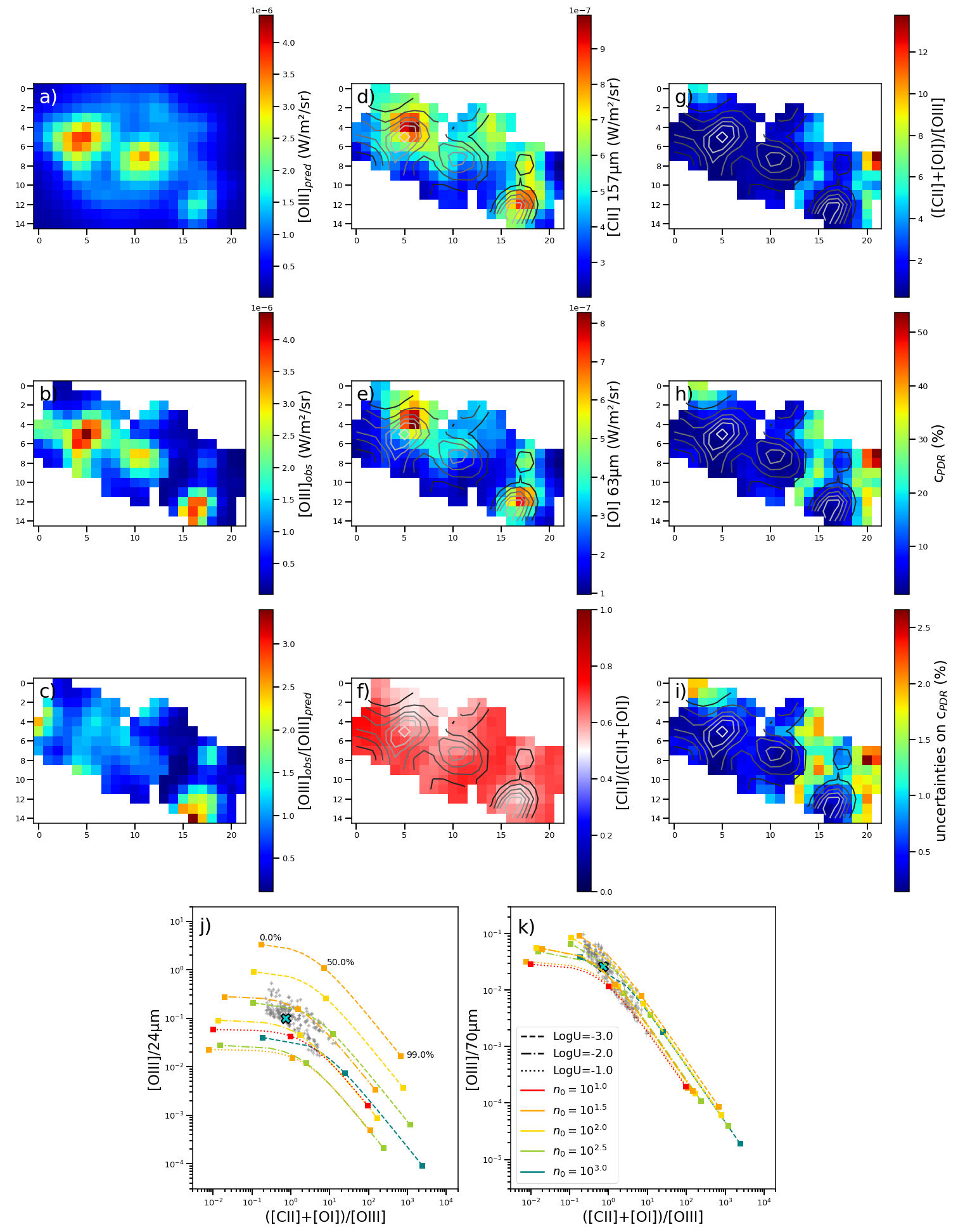}
\caption{Similar to Figure \ref{fig-app_30Dor}, but for the region N159.}
\label{fig-app_N159}
\end{center}
\end{figure*}

\newpage

\begin{figure*}[h!]
\begin{center}
\includegraphics[scale=0.3]{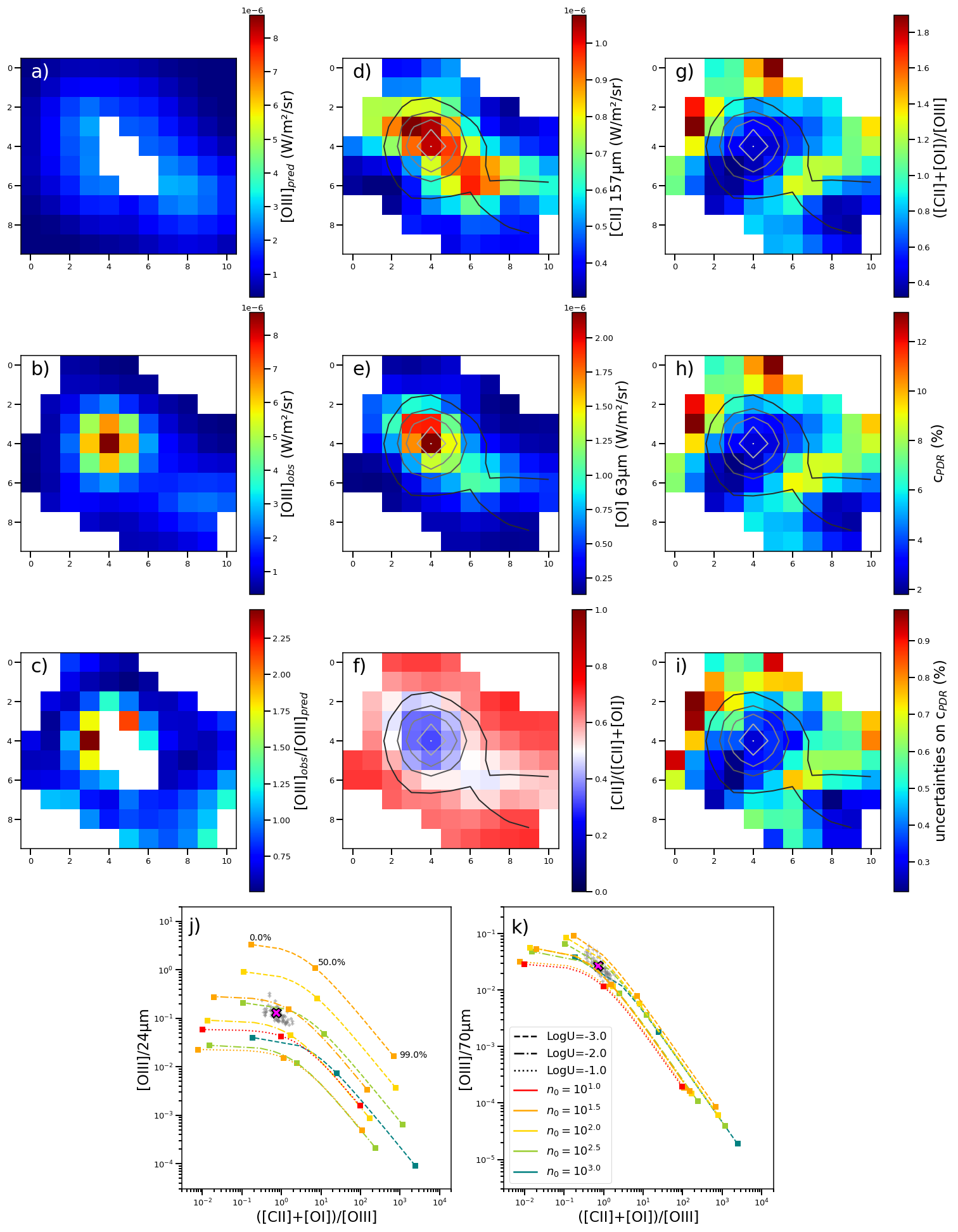}
\caption{Similar to Figure \ref{fig-app_30Dor}, but for the region N160.}
\label{fig-app_N160}
\end{center}
\end{figure*}

\newpage

\begin{figure*}[h!]
\begin{center}
\includegraphics[scale=0.3]{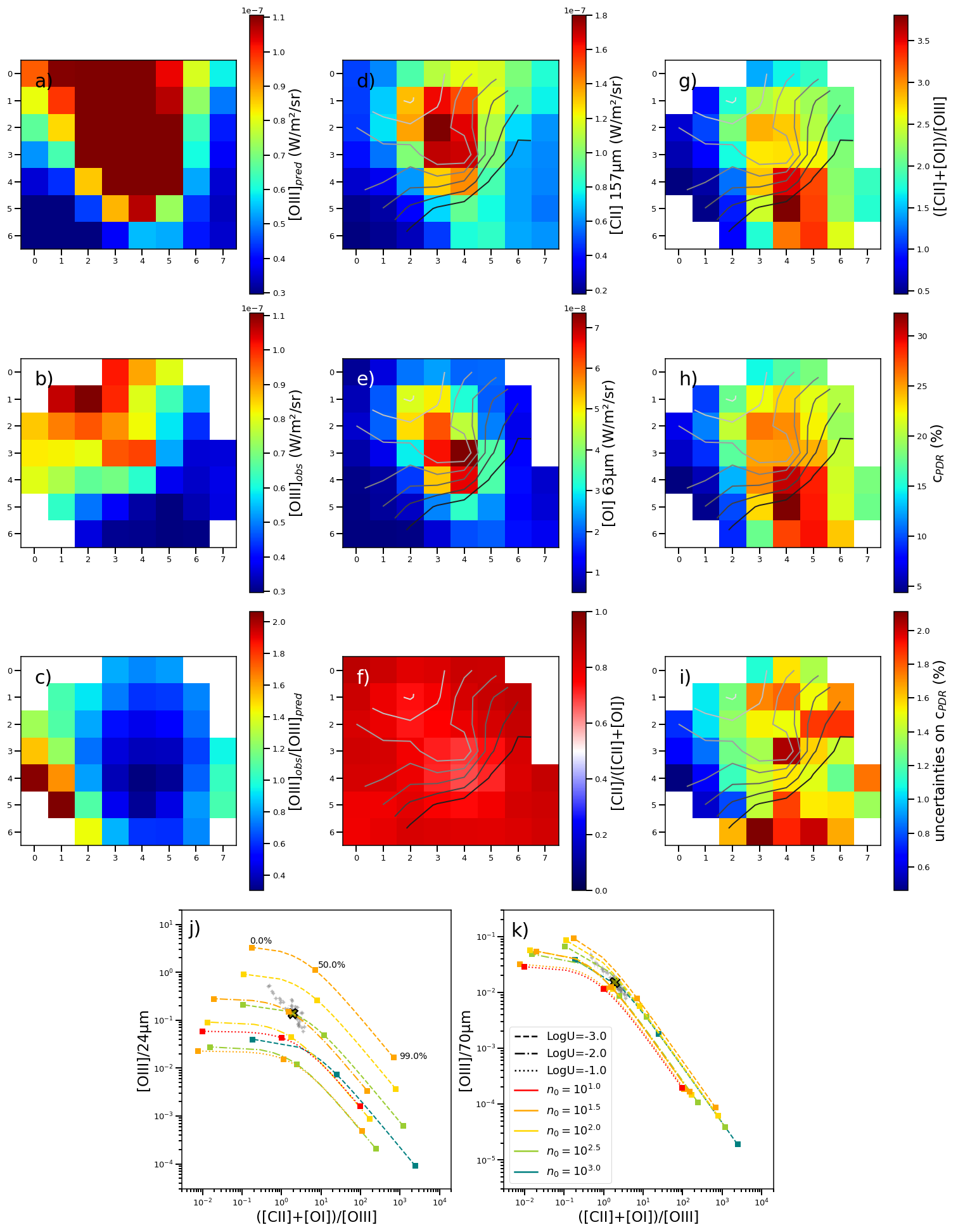}
\caption{Similar to Figure \ref{fig-app_30Dor}, but for the region N180.}
\label{fig-app_N180}
\end{center}
\end{figure*}

\newpage

\begin{figure*}[h!]
\begin{center}
\includegraphics[scale=0.3]{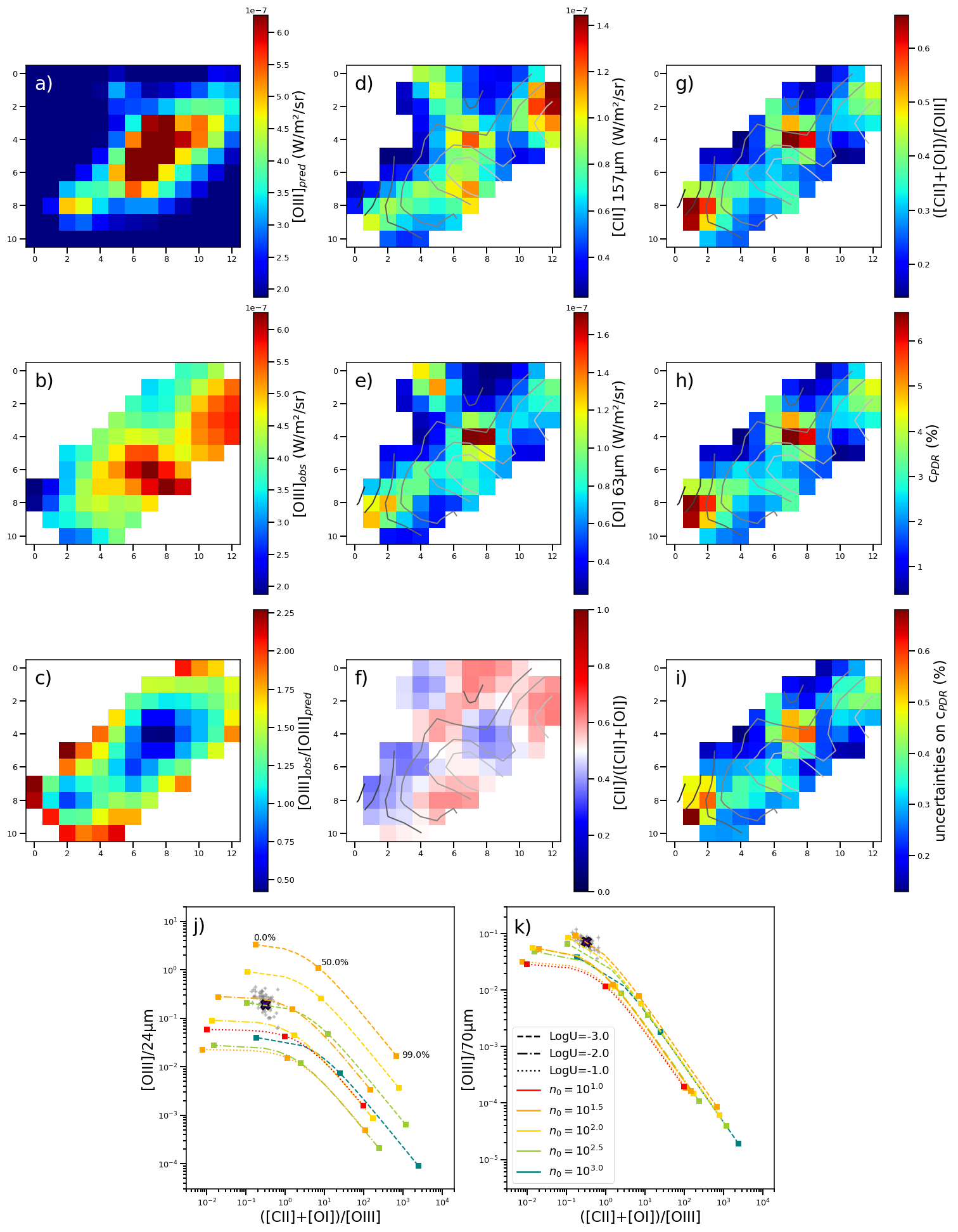}
\caption{Similar to Figure \ref{fig-app_30Dor}, but for the region N66.}
\label{fig-app_N66}
\end{center}
\end{figure*}

\end{appendix}

\end{document}